\documentclass[preprint,10pt]{elsarticle}
\usepackage{amsfonts,amsmath,amssymb,epsfig,graphicx,algorithm,subfig}
\usepackage{booktabs, wrapfig}

\makeatletter
    \newcommand{\Rmn}[1]{\it{\expandafter\@slowromancap\romannumeral #1@}}
\makeatother

\journal{Journal of Computational Physics}

\begin{document}
\begin{frontmatter}

\title{Least Squares Shadowing sensitivity analysis of chaotic limit cycle oscillations}

\author[mit]{Qiqi Wang\corref{cor}}
\ead{qiqi@mit.edu}

\author[mit]{Rui Hu}
\ead{hurui@mit.edu}

\author[mit]{Patrick Blonigan}
\ead{blonigan@mit.edu}

\cortext[cor]{Corresponding author.}
\address[mit]{Aeronautics and Astronautics,
MIT, 77 Mass Ave, Cambridge, MA 02139, USA}

\begin{abstract}
The adjoint method, among other sensitivity analysis
methods, can fail in chaotic dynamical systems.
The result from these methods can be too large, often
by orders of magnitude, when the result is the derivative of
a long time averaged quantity.  This failure is known to be caused
by ill-conditioned initial value problems.
This paper overcomes this failure by replacing the initial value problem
with the well-conditioned ``least squares shadowing (LSS) problem''.
The LSS problem is then linearized in our sensitivity analysis algorithm,
which computes a derivative that converges to the derivative of
the infinitely long time average.  We demonstrate our algorithm
in several dynamical systems exhibiting both periodic and chaotic oscillations.
\end{abstract}

\begin{keyword}
    Sensitivity analysis, linear response, adjoint equation,
    unsteady adjoint, chaos, statistics, climate, least squares shadowing
\end{keyword}

\end{frontmatter}

\section{Introduction}
\label{s:intro}

As more scientists and engineers use computer simulations,
some begins to harness the versatile power of sensitivity analysis.
It helps them engineer products \cite{Jameson1988,adjoint2},
control processes and systems \cite{adjoint1,bewley01},
solve inverse problems \cite{seismic_adjoint},
estimate simulation errors \cite{Becker_2001_An_Optimal_Control_Approach,
Giles2002, error2010, fidkowski2011review},
assimilate measurement data \cite{QJ:QJ49711750206,weather4}
and quantify uncertainties \cite{qiqiwang_thesis}.

Sensitivity analysis computes the derivative of outputs to inputs of a
simulation.  Conventional methods, including the tangent and the adjoint
method, are introduced in Section
\ref{s:conventional}.  These methods, however, fails when the dynamical
system is chaotic and the outputs are long time averaged
quantities.   They compute derivatives that are
orders of magnitude too large, and that grow exponentially larger
as the simulation runs longer.  What causes this failure is
the ``butterfly effect'' -- sensitivity of chaotic initial value
problems.  This diagnosis is first published by Lea et al \cite{leaclimate}, 
and explained in Section \ref{s:breakdown}.

Many researchers have become interested in overcoming this failure,
a challenge in both dynamical systems and numerical methods.
They have recently developed a few methods for
computing \emph{useful} derivatives of long time averaged outputs
in chaotic dynamical systems.  Lea et al pioneered the ensemble
adjoint method \cite{leaclimate,eyinkclimate}, which
applies the adjoint method to many random trajectories,
then averages the computed derivatives.
Nevertheless, they need impractically many trajectories,
making the method costly even for small
dynamical systems such as the Lorenz system.
Thuburn introduced an approach that solves the adjoint of the
Fokker-Planck equation, which governs a probability distribution in the phase
space \cite{QJ:QJ200513160505}.  However, this approach assumes the
probability distribution to be smooth, a property often achieved by
by adding dissipation to the Fokker Planck equation, causing
error in the result.  

In addition, researchers have adopted the Fluctuation-Dissipation Theorem
for sensitivity analysis \cite{0951-7715-20-12-004}.  This approach have
several variants.  Different variants, however, has different limitations.
Some assume the dynamical system to have an equilibrium distribution
similar to the Gaussian distribution, an
assumption often violated in dissipative dynamical systems.
Other variants nonparametrically estimate the equilibrium distribution
\cite{cooper2011climate}, but add artificial
noise to the dynamical system to ensure its smoothness.
The first author recently used Lyapunov eigenvector decomposition for
sensitivity analysis \cite{wangLorenz}.
However, this decomposition
requires high computational cost when the dynamical system has many positive
Lyapunov exponents.  Despite these new methods, nobody has applied
sensitivity analysis to long time averaged outputs in turbulent flows,
or other large, dissipative and chaotic systems.

This paper presents the \emph{Least Squares Shadowing method},
a new method for computing derivatives
of long time averaged outputs in chaos.
The method linearizes the \emph{least
squares shadowing problem}, a constrained least squares problem
defined in Section \ref{s:shadowing}.
It then solves the linearized problem with a numerical method
described in Section \ref{s:numerical}.
Demonstrated with three application
in Sections \ref{s:vdp}, \ref{s:lorenz} and \ref{s:aeroelastic},
the method is concluded in Section \ref{s:conclude} to be
potentially useful in large chaotic dynamical systems.

\section{Conventional method for sensitivity analysis}
\label{s:conventional}

In sensitivity analysis, an output $J$ depends on an input $s$
via a simulation, which solves an ordinary differential equation
\begin{equation} \label{ode}
\frac{du}{dt} = f(u, s)
\end{equation}
starting from an initial condition
\begin{equation} \label{odeiv} u|_{t=0} = u_0(s) \;,\end{equation}
where the input $s$ can
represent control variables, design variables, and
uncertain parameters.  This initial value problem
(\ref{ode}-\ref{odeiv})
determines a solution $u_{iv}(t;s)$ that depends on time and the input.

An output $J(u,s)$ is a function of the solution and the input.
It can also be viewed as a function of time and the input by
substituting the solution $u_{iv}(t;s)$.
The time averaged output,
\begin{equation} \label{finiteobj}
\overline{J}^{(T)}_{iv}(s) := \frac1T \int_0^T J(u_{iv}(t;s), s) \,
dt\;, \end{equation}
then depends only on the input $s$.
Its derivative to $s$ can be computed by
the conventional tangent method of sensitivity
analysis \cite{brysonho}.

The conventional tangent method first solves the linearized
governing equation, also known as the \emph{tangent equation},
\begin{equation} \label{linearized}
  \frac{dv}{dt} = \frac{\partial f(u_{iv},s)}{\partial u} v
            + \frac{\partial f(u_{iv},s)}{\partial s}
\end{equation}
with the linearized initial condition
\begin{equation}
            \quad v|_{t=0} = \frac{d u_0}{ds}\;.
\end{equation}
The solution $v_{iv}(t;s)$ indicates how a small change in $s$ alters the
solution to the initial value problem $u_{iv}(t;s)$:
\begin{equation}
  v_{iv}(t;s) = \frac{\partial u_{iv}(t;s)}{\partial s}
\end{equation}
This solution is then used
to compute the derivative of $\overline{J}_{iv}^{(T)}(s)$:
\begin{equation}\label{overlineJder} \frac{d \overline{J}_{iv}^{(T)}}{d s}
 = \frac1T \int_0^T \left(\frac{\partial J(u_{iv},s)}{\partial u} v_{iv}
                + \frac{d J(u_{iv},s)}{d s} \right) dt \end{equation}

This method can be transformed into the conventional adjoint method
\cite{brysonho}, which computes the derivative of
one objective function to many inputs simultaneously.
This advantage makes the adjoint method popular in optimal control,
inverse problems and data assimilation applications.

\section{Failure of the conventional method for time averaged outputs
in chaos}
\label{s:breakdown}

The conventional method fails when the simulation
(\ref{ode}) is chaotic, and the output (\ref{finiteobj}) is
averaged over a long time $T$.  A chaotic dynamical
system is sensitive to its initial condition, causing the solution to the
linearized initial value problem (\ref{linearized}) to grow at a
rate of $e^{\lambda t}$, where $\lambda>0$
is the maximal Lyapunov exponent of the dynamical system.
This exponential growth makes $v_{iv}(t;s)$ large 
unless $t$ is small.  When substituted into Equation
(\ref{overlineJder}), we expect a large
$\frac{d \overline{J}_{iv}^{(T)}}{d s}$ unless $T$ is small.

The value of $\frac{d \overline{J}_{iv}^{(T)}}{d s}$ can exceed
$10^{100}$ time of what scientists had expected.
Lea et al. \cite{leaclimate} documented this in the Lorenz system,
which models heat convecting from a warm horizontal surface to a cooler
one placed above it.  Their temperature difference, described by the Rayleigh
number, affects how fast the heat convects; it is therefore chosen by
Lea at al as the input $s$.
The heat convection rate is chosen as the output $J$;
its time average should increase with $s$ at a ratio of about 1.
\footnote{In Lea et al.'s original paper,
the Rayleigh number is denoted
as $\rho$ and the convective heat transfer rate is denoted as $z$.
These notations are conventional in Lorenz system literature.
But in this paper, we denote the Rayleigh number as $s$ and
the heat transfer rate as $J$, so that we are consistent with the
general notation of input and output.}

\begin{figure}[htb!] \centering
\subfloat[$\overline{J}_{iv}^{(T)}(s)$ for $T=2.26$
]{\includegraphics[width=2.2in]{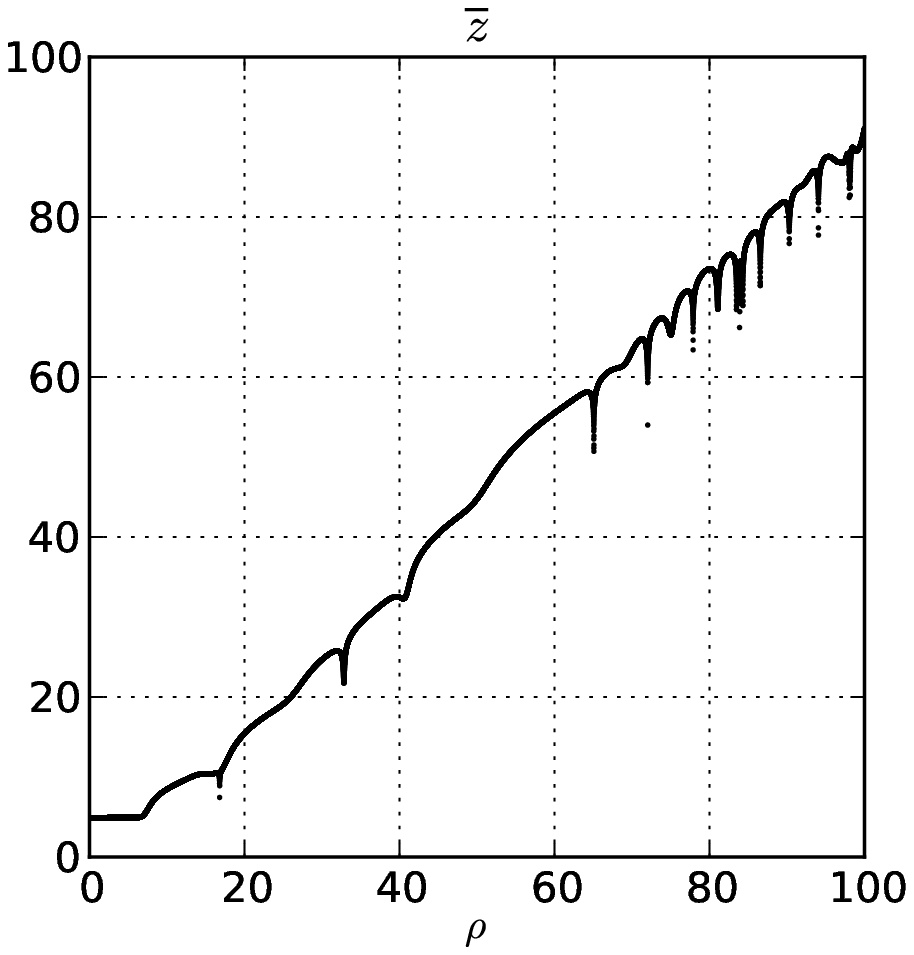}}
    \hspace{0.1in}
\subfloat[$\left|\frac{d\overline{J}_{iv}^{(T)}}{ds}\right|$ for $T=2.26$
]{\includegraphics[width=2.2in]{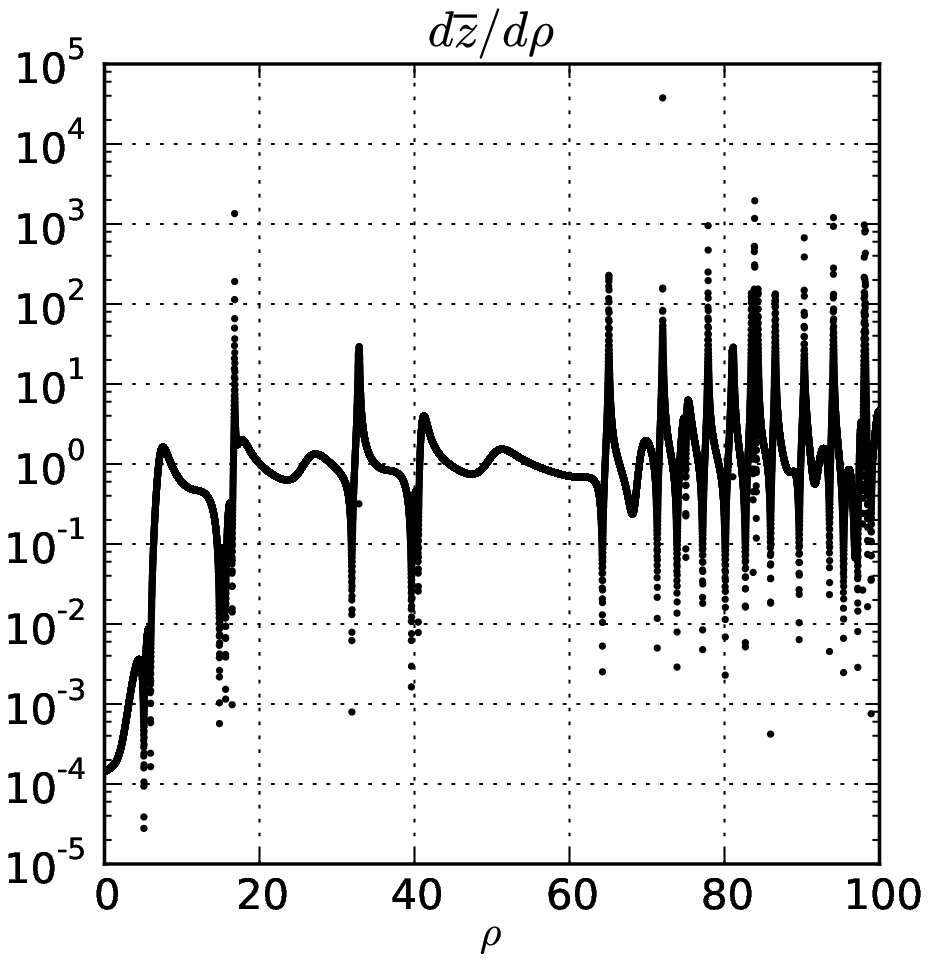}}\\
\subfloat[$\overline{J}_{iv}^{(T)}(s)$ for $T=131.4$
]{\includegraphics[width=2.2in]{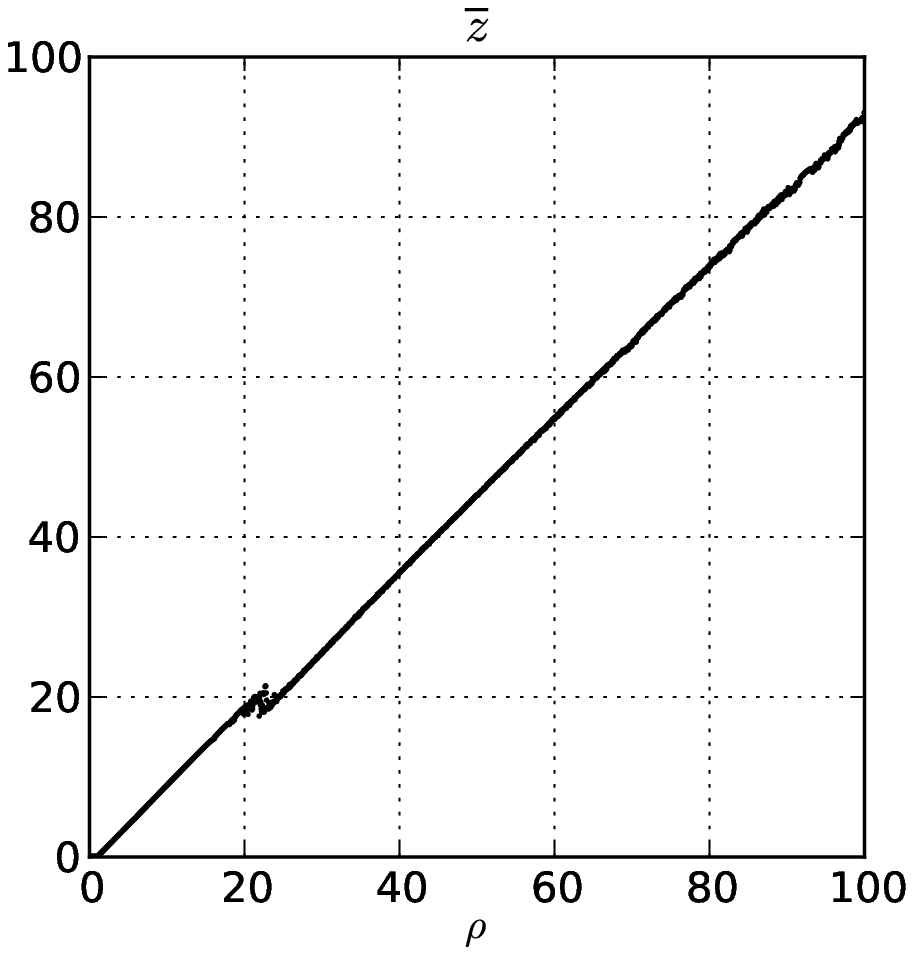}}
    \hspace{0.1in}
\subfloat[$\left|\frac{\overline{J}_{iv}^{(T)}}{d s}\right|$ for $T=131.4$
]{\includegraphics[width=2.2in]{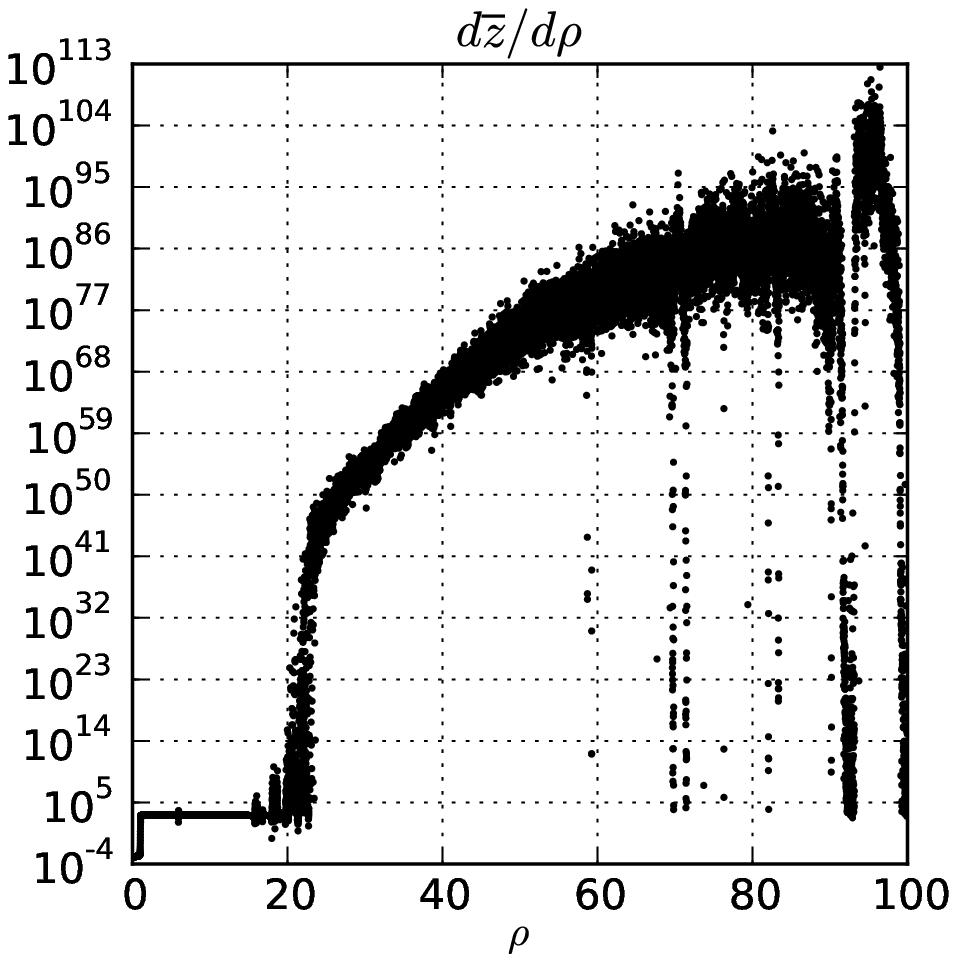}}
\caption{Plots created following the procedure in Lea et al\cite{leaclimate}
(permission granted).  Left: time averaged output $\overline{J}_{iv}^{(T)}(s)$
plotted against the input $s$.  Right: the derivative of
the time averaged output with respect to $s$.  Note the order of
magnitude of the $y$-axes.}
\label{f:lea01}
\end{figure}

Lea et al. considered a range of input $s$
and several values of the averaging length $T$.  At each $s$ and $T$,
they simulated the Lorenz system and computed
$\overline{J}^{(T)}_{iv}(s)$.  They then computed the derivative
$\frac{d \overline{J}_{iv}^{(T)}}{d s}$ using the conventional
adjoint sensitivity analysis method.  When $T$ is large, they
found the derivative of
$d \overline{J}_{iv}^{(T)}$ orders of magnitude larger
than its expected slope of about 1.
By repeating Lea et al.'s procedure, we found that the astronomical values of
$\frac{d \overline{J}_{iv}^{(T)}}{ds}$, plotted in Figure \ref{f:lea01}, are
insensitive to how Equations (\ref{ode}-\ref{overlineJder}) are discretized.

The computed derivative $\frac{d \overline{J}_{iv}^{(T)}}{d s}$
is too large to be useful.
The derivative is useful in approximating the slope of the function,
$\frac{\overline{J}_{iv}^{(T)}(s+\delta s) -
\overline{J}_{iv}^{(T)}(s)}{\delta s}$.
The better it approximates this slope, and over a larger interval
size $\delta s$, the more useful it is.
If the derivative is as large as $10^{50}$, the function must
have a correspondingly steep slope when plotted against $s$,
but only so monotonically over intervals smaller than $10^{-50}$.
The derivative can approximate the slope of the function well
only within these impractically tiny intervals -- computers
cannot even represent an interval of
$[1, 1+10^{-16}]$ in double precision.
For approximating the slope of the function over a practical interval
$[s,s+\delta s]$, the derivative is useless.

This failure happens not only to the Lorenz system, but to other
chaotic dynamical systems such as chaotic
fluid flows \cite{wanggao}.  It is caused by the sensitivity of chaos.
Popularly known as the ``butterfly effect'', this sensitivity
makes the finite time average $\overline{J}_{iv}^{(T)}$ ill-behaved,
its derivative with respect to $s$ fluctuating wildly.
A small change in $s$ almost always causes a large change in the solution
$u_{iv}$, thus a large change in the tangent solution
$v_{iv}$, and thus a large change in the derivative
$\frac{d \overline{J}_{iv}^{(T)}}{d s}$.
As an $s$ increases to $s+\delta s$,
the derivative can vary over a wide range of positive and negative
values.  These derivative values, by the fundamental
theorem of calculus, must average to the slope of the function
\begin{equation} \label{ftc}
    \mbox{slope}:=
\frac{\overline{J}_{iv}^{(T)}(s+\delta s) -
\overline{J}_{iv}^{(T)}(s)}{\delta s}
=
\frac{1}{\delta s} \int_{s}^{s+\delta s}
\frac{d \overline{J}_{iv}^{(T)}}{ds} ds'\;,
\end{equation}
but because the derivative fluctuates rapidly and wildly between extreme
values of either sign, at almost any point within $[s,s+\delta s]$, the
derivative is much larger in magnitude than the slope of the function
over $[s,s+\delta s]$.

How sensitive a solution $u$ is to its
input $s$ can be quantified by the \emph{condition number}, defined as
$\|du/ds\|$.  We call a problem ill-conditioned if it has a large
condition number, or well-conditioned if it has a small one.
A chaotic initial value problem has a condition number on the order of
$e^{\lambda T}$, where $\lambda$ is the maximal Lyapunov exponent.
Even moderately long simulations can be ill-conditioned, causing sensitivity
analysis to fail.  To overcome this failure,
we must substitute the initial value problem with a
well-conditioned one.

\section{Sensitivity analysis via Least Squares Shadowing}
\label{s:shadowing}

\subsection{The nonlinear Least Squares Shadowing (LSS) problem}

The initial condition of a simulation can be relaxed
if the following assumptions hold:
\begin{enumerate}
\item {\bf We are interested in infinite time averaged outputs.}
When scientists and engineers compute a long time averaged output, they
often intend to approximate the limit
\begin{equation} \label{infiniteobj}
\overline{J}^{(\infty)}(s)
:= \lim_{T\to\infty}\frac1T \int_0^T J(u(t;s), s) \, dt\;.
\end{equation}
We assume that these infinite time averaged outputs, and functions
thereof, are the only outputs of interest.
\item {\bf The dynamical system is \emph{ergodic}.}
An ergodic dynamical system behaves the same over long
time, independent of its initial condition.
Specifically, the initial condition does not affect an
infinite time averaged outputs defined above.
\end{enumerate}

Under these two assumptions, we can approximate the outputs
using a long solution of the governing equation, regardless
of where the solution starts.  We
replace initial condition with a criterion that makes the problem
better-conditioned.  Among all trajectories that satisfy the
governing equation, we chose
one that is closest to a pre-specified reference trajectory $u_r$
in the following metric:
\begin{equation} \label{lsq0} \begin{split}
 &  \underset{\tau, u}{\mbox{minimize }} \frac1T\int_{0}^{T} \left(\Big\|u(\tau(t)) - u_r(t)\Big\|^2
   + \alpha^2 \left(\frac{d\tau}{dt} - 1\right)^2\right) dt \;,\\
 & \quad\mbox{such that} \qquad\frac{du}{dt} = f(u,s)\;.
\end{split}\end{equation}
We choose the reference trajectory $u_r(t)$ to be a solution to the governing
equation at a different $s$, set the constant $\alpha$ so that the two
terms in the integral have similar magnitude, then
minimize this metric among all trajectories $u(t)$ and all
monotonically increasing time transformations $\tau(t)$.

We call this constrained minimization problem (\ref{lsq0}) the Least
Squares Shadowing (LSS) problem.  We denote its solution
as $u_{lss}^{(T)}(t;s)$ and $\tau_{lss}^{(T)}(t;s)$.
They are
a solution of the governing equation and a time transformation that makes 
this solution close to $u_r$.
Because $u_{lss}^{(T)}(t;s)$
satisfies the governing equation, we use it to approximate
\begin{equation} \label{lssobj}
\overline{J}^{(\infty)}(s)
\approx \overline{J}^{(T)}_{lss}(s)
:= \frac1{\tau(T)-\tau(0)} \int_{\tau(0)}^{\tau(T)}
J(u^{(T)}_{lss}(t;s), s) \, dt\;.
\end{equation}
with sufficiently large $T$.

\subsection{Well-conditioning of the Least Squares Shadowing (LSS) problem}

An initial value problem of chaos is ill-conditioned, causing failure to
conventional sensitivity analysis methods, a failure we now overcome
by switching to the LSS problem, a well-conditioned problem
whose solution is less sensitive to perturbations in the parameter
value, and whose long time averages have useful derivatives.

\begin{figure}[htb!]
\subfloat[$J(u_{iv}(t;s), s)$]
{\includegraphics[width=0.48\textwidth]{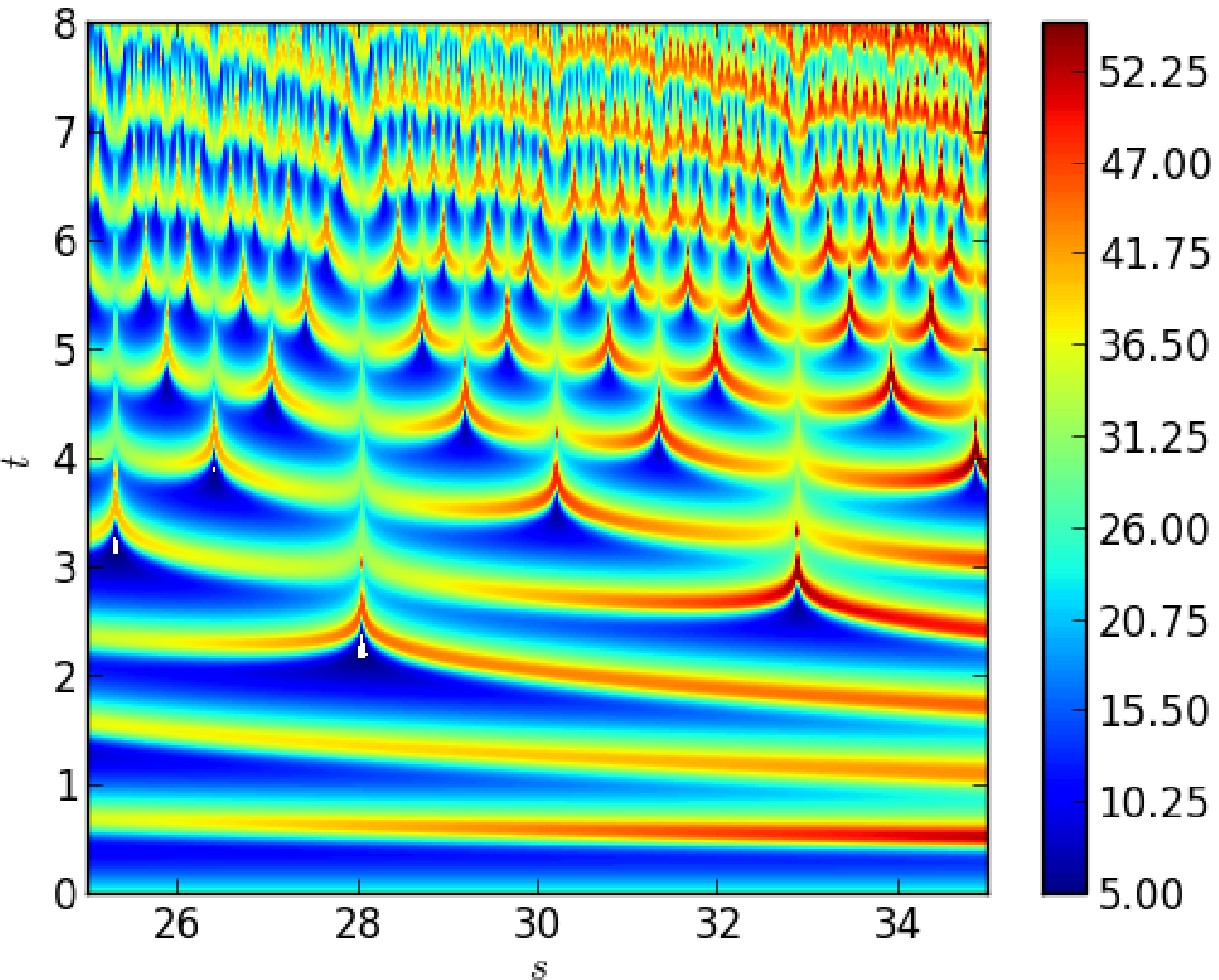} }
\subfloat[$J(u_{lss}^{(T)}(t;s),s)$ and $\tau_{lss}^{(T)}(t;s)$.]
{\includegraphics[width=0.48\textwidth]{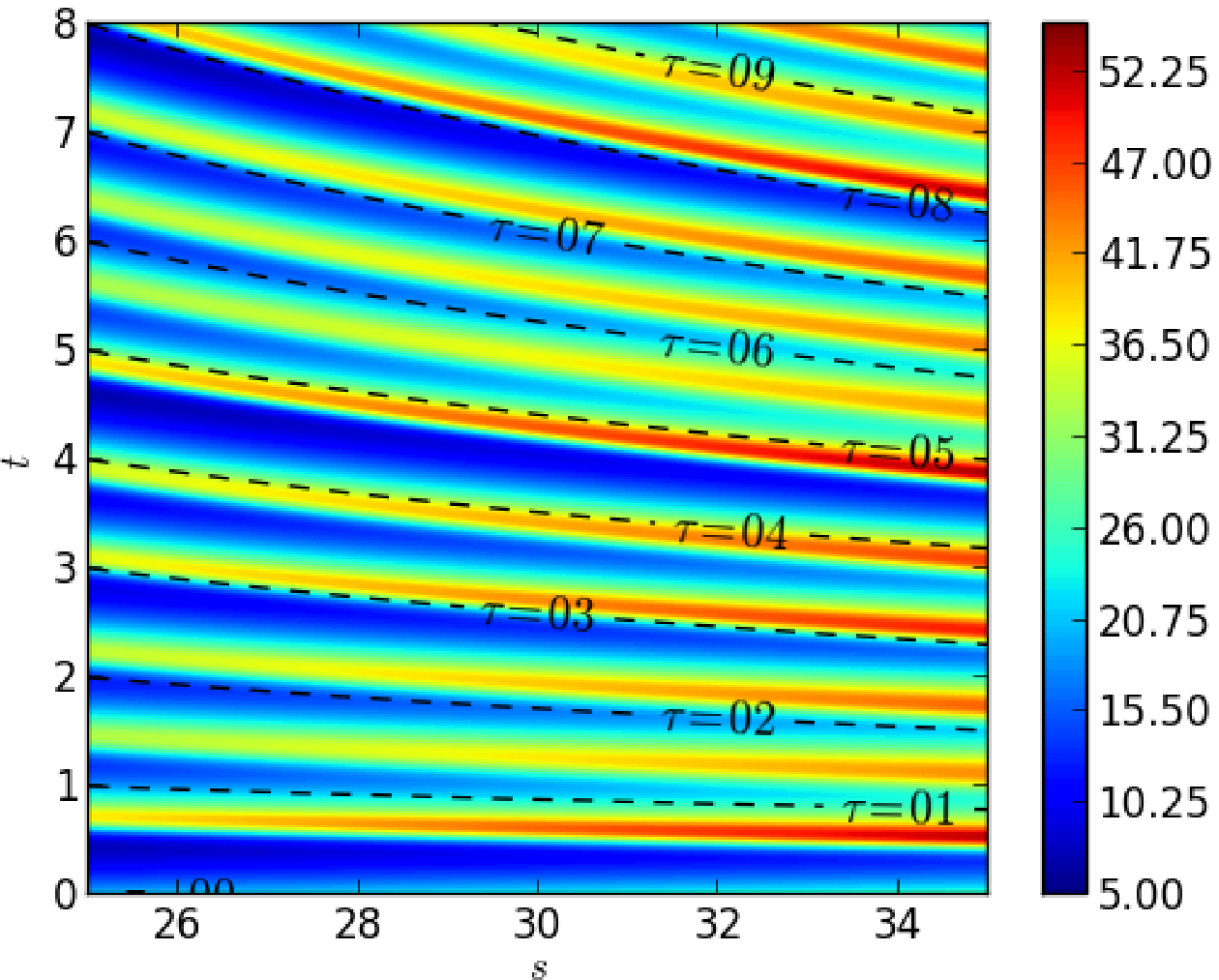}}
\hspace{0.02\textwidth}
\caption{Time dependent rate of heat transfer in the Lorenz system
with varying Rayleigh number $s$.  This output is computed by solving
initial value problems in the left plot,
and by solving LSS problems in the right plot.
Each vertical slice represents the time dependent output
at an $s$ value.}
\label{f:ivplsp}
\end{figure}

\begin{wrapfigure}{r}{0.5\textwidth}\vspace{-0.04\textwidth}
\includegraphics[width=0.5\textwidth]{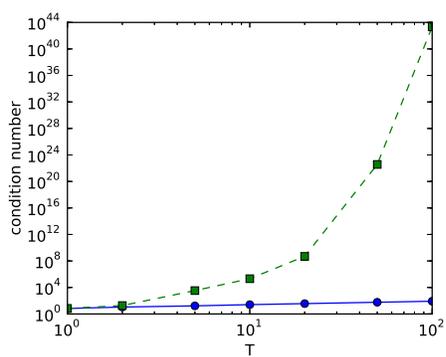}\vspace{-0.04\textwidth}
\caption{The condition number increases rapidly in an initial value
problem (dashed line with squares),
but stays relatively constant in an LSS problem (solid line with
circles).}
\label{f:cond}
\end{wrapfigure}

Figure \ref{f:ivplsp} visualizes how sensitive the initial value problem
is, whereas how robust the LSS problem is\footnote{In Figure \ref{f:ivplsp}(b),
we solve a single initial
value problem at $s=25$, followed by a sequence of Least squares problems
at increasing values of $s$, each using the previous
solution as its reference trajectory $u_r$.}.
The initial value problem produces
solutions that grows more sensitive to the input $s$
as time advances.  Its condition number
grows exponentially as the trajectory length increases.
The LSS problem produces solutions that
gradually depend on $s$.  As shown in Figure \ref{f:cond},
it stays well-conditioned regardless of how
long the trajectory is.

The LSS problem is well-conditioned, a result not
only observed in the Lorenz system, but also 
derives from the \emph{shadowing lemma}\cite{pilyugin1999shadowing}.
It guarantees that a trajectory of the governing equation
exists in the proximity of any ``$\delta$-pseudo
trajectory'', defined as an approximate solution that satisfies
the governing equation to $\delta$-precision.  The lemma assumes
a set of properties known as \emph{uniform
hyperbolicity}\cite{kuznetsov2012hyperbolic, CambridgeJournals:1825204},
and states that \emph{for any $\epsilon>0$,
there exists $\delta$, such that for all $\delta$-pseudo
trajectory $u_r$ of any length, there exists a true trajectory $u$
within $\epsilon$ distance from $u_r$, in the same distance metric
used in Equation (\ref{lsq0})}.
If $u_r$ is a true trajectory at input value
$s$, and thereby a $\delta = \sup\frac{\partial f(u;s)}{\partial s}\;\delta s$
-pseudo-trajectory at input value $s+\delta s$,
then the shadowing lemma predicts the LSS
solution $u_{lss}$ to be within $\epsilon$ distance from $u_r$.
Perturbing $s$ slightly makes $u_{lss}$ slightly
different from $u_r$, indicating a well-conditioned problem
regardless of how long the trajectory is.

Because the LSS problem is well-conditioned, its
time averaged output $\overline{J}_{lss}^{(T)}(s)$ has a useful
derivative.  This LSS derivative $\frac{d \overline{J}_{lss}^{(T)}}{ds}$
can be computed by solving a
linearized LSS problem (detailed in Section \ref{s:method}).
Because of its well-conditioning, perturbing the input between $s$ and
$s+\delta s$ causes a small difference in its solution,
and therefore a small difference in the LSS derivative.
This, and the fundamental theorem of calculus
\begin{equation} \label{ftc1}
\frac{1}{\delta s} \int_{s}^{s+\delta s}
\frac{d \overline{J}_{lss}^{(T)}}{ds} ds
= 
\frac{\overline{J}_{lss}^{(T)}(s+\delta s) -
\overline{J}_{lss}^{(T)}(s)}{\delta s}\;,
\end{equation}
make the LSS derivative at any $s\in[s,s+\delta s]$ a useful
approximation to the slope.

As $T\to\infty$, this slope
converges to the slope of the infinite time average
$\overline{J}^{(\infty)}$, and the LSS derivative converges to the
derivative of this infinite time average.  Such derivative exists
not only as a derivative of the limit (\ref{infiniteobj})
\cite{springerlink10,CambridgeJournals:1825204},
but also as a limit of the LSS derivative as $T\to\infty$.
The limit and the derivative commute because the slope of
$\overline{J}^{(\infty)}$ between $s$ and $s+\delta s$
uniformly converges to its derivative as $\delta s$ vanishes -- a proven
result made possible by the well-conditioned LSS problem \cite{lsstheory}.

\subsection{Computing derivative from linearized Least Squares Shadowing
(LSS) solution}
\label{s:method}

The linearized LSS problem derives from the nonlinear 
problem (\ref{lsq0}).  We choose a
reference trajectory $u_r$ that satisfies the governing equation at an
input value $s$, then perturb $s$ by an infinitesimal
$\delta s$.  By ignoring $O(\delta s^2)$ terms in Taylor
expansions, we obtain
\begin{equation} \label{lsq2} \begin{split}
 &  \underset{\eta, v}{\mbox{minimize }} \frac1T\int_{0}^{T}
   \left(\|v\|^2 + \alpha^2 \eta^2\right)dt\;, \quad\mbox{such that}\\
 & \frac{dv}{dt} = \frac{\partial f}{\partial u} v
            + \frac{\partial f}{\partial s} + \eta f(u_r,s)\;,
\end{split}\end{equation}
where $v(t)$ and $\eta(t)$ are the solution of this linearized LSS
problem.  They relate to the solution of the nonlinear
problem $\tau^{(T)}_{lss}$ and $u^{(T)}_{lss}$ via
\begin{equation}
v(t) =
\frac{d}{ds}\bigg(u^{(T)}_{lss}\left(\tau^{(T)}_{lss}(t;s);s\right)\bigg)
\;,\quad
\eta(t) = \frac{d}{ds} \frac{d\tau^{(T)}_{lss}(t;s)}{dt}\;.
\end{equation}
The linearization is detailed in the Appendix.
We also linearize the time averaged output
$\overline{J}_{lss}^{(T)}$ as defined in Equation (\ref{lssobj}),
and obtain a formula for computing the desired derivative from the
solution of the linearized LSS problem
\begin{equation} \label{climatesens}
\frac{d\langle J\rangle}{ds} \approx \frac{
   \displaystyle\int_{0}^{T}\left(
   \frac{\partial J}{\partial u}v
 + \frac{\partial J}{\partial s}
 + \eta \left( J - \overline{J}\:\right)\right)dt}{T} \;,
 \quad\mbox{where}\quad
 \overline{J}=\frac{\displaystyle\int_{0}^{T} J\,dt}{T} 
\end{equation}
This linearization is also derived in the Appendix.

\section{Numerical solution of the Least Squares Shadowing (LSS) problem}
\label{s:numerical}
The linearized LSS
problem (\ref{lsq2}) can be solved with two numerical approaches.
One approach, detailed in Subsection \ref{s:disckkt},
first discretizes Problem (\ref{lsq2}), then derive
from the discretized minimization problem its optimality condition,
a system of linear equations that are finally solved to obtain the solution
$v$ and $\eta$.
The other approach, detailed in Subsection \ref{s:kktdisc},
applies variational calculus to 
Problem (\ref{lsq2}) to derive its variational optimality
condition, a system of linear differential equations that are then discretized
and solved to obtain $v$ and $\eta$.
Both approaches can lead to the same linear
system, whose solution method is described in Subsection
\ref{s:solvekkt}.  Section \ref{s:algorithm} provides a short summary
of the numerical procedure.  The algorithm admits 
an adjoint counterpart, described in 
Subsection \ref{s:adjoint}, that can compute derivatives
to many parameters simultaneously.

\subsection{Derivation of the linear system via
the discrete optimization approach}
\label{s:disckkt}
We first convert Problem (\ref{lsq2})
from a variational minimization problem to a finite dimensional
minimization problem.  By dividing the time domain $[0,T]$ into
$m=T/\Delta t$ uniform time steps\footnote{
$\Delta t$ is chosen to be uniform for all time steps because it simplifies
the notation.  The algorithm can be extended to nonuniform $\Delta t$,
as implemented in the lssode package\cite{lssode_sftw}.
}, denoting $u_{i+\frac12} = u_r\left(\left(i + \frac12\right)\Delta t\right),
v_{i+\frac12} = v\left(\left(i + \frac12\right)\Delta t\right),
i=0,\ldots,m-1$ and $\eta_{i}=\eta(i \Delta t), i=1,\ldots,m-1$,
and approximating the time derivatives of $u$ and $v$ via the
trapezoidal rule\footnote{
We choose the trapezoidal rule because it is single-step and
second-order accurate.  Other time discretization can be used, though
the resulting system will be either more complex or less accurate.
}, we discretize the linearized LSS
problem (\ref{lsq2}) into
\begin{equation} \label{dlsq} \begin{split}
    &\underset{v_i, \eta_i}{\mbox{minimize }} \sum_{i=0}^{m-1} \frac{\|v_{i+\frac12}\|_2^2}{2}
 + \alpha^2\sum_{i=1}^{m-1} \frac{\eta_{i}^2}{2}\;,\qquad \mbox{such that}\\
   & E_i v_{i-\frac12} + f_i \eta_i + G_i v_{i+\frac12} = b_i\;,\quad 1\le i< m
\end{split}\end{equation}
where
\begin{equation} \label{efgdef} \begin{split}
   E_i &= -\frac{I}{\Delta t} -
          \frac{\partial f}{\partial u}(u_{i-\frac12}, s)\;, \\
   f_i &= \frac{u_{i+\frac12} - u_{i-\frac12}}{\Delta t}\;,\\
   G_i &= \frac{I}{\Delta t} -
          \frac{\partial f}{\partial u}(u_{i+\frac12}, s)\;. \\
   b_i &= \frac12 \left(\frac{\partial f(u_{i-\frac12}, s)}{\partial s}
                       +\frac{\partial f(u_{i+\frac12}, s)}{\partial s}
                  \right)\;,\\
\end{split}\end{equation}
This linear-constrained least-squares
problem has an optimality condition that forms
the following KKT system\cite{boyd2004convex}
\begin{equation} \label{kkt} \addtolength{\arraycolsep}{-1mm}
{\scriptstyle
\left[\begin{array}{cccccccc|cccc}
I   &   &    &   &    &        &     &   &    E_1^T &        &        &     \\
    &\alpha^2 &    &   &    &        &     &   &    f_1^T &        &        &     \\
    & & I    &   &    &        &     &   &    G_1^T & E_2^T  &        &     \\
    & &      & \alpha^2 &    &        &     &   &          & f_2^T  &        &     \\
    & &      &   & I &        &     &   &          & G_2^T  &        &     \\
    & &      &   &    & \ddots &     &   &          &        & \ddots & E_{\scriptscriptstyle m-1}^T \\
    & &      &   &    &        & \alpha^2 &   &          &        & & f_{\scriptscriptstyle m-1}^T \\
    & &      &   &    &        &     & I &          &        & & G_{\scriptscriptstyle m-1}^T \\
\midrule
E_1 & f_1 & G_1    &        &        &   &    &       &        &        &     \\
    & & E_2 & f_2 & G_2    &        &    &   &       &        &        &     \\
    & &  &    & \ddots & \ddots &      & &     &        &        &     \\
    & &  &   &        & E_{\scriptscriptstyle m-1}& f_{\scriptscriptstyle m-1} & G_{\scriptscriptstyle m-1}    &       &        &        &     \\
\end{array}\right]
\left[\begin{array}{c}
v_{\frac12} \\ \eta_1 \\ v_{1+\frac12} \\ \eta_2 \\ v_{2+\frac12}
\\ \vdots \\ \eta_{\scriptscriptstyle m-1} \\ v_{m-\frac12} \\
\midrule w_1 \\ w_2 \\ \vdots \\ w_{\scriptscriptstyle m-1}
\end{array}\right]
=
\left[\begin{array}{c}
  0 \\ 0 \\ 0 \\ 0 \\ 0 \\ \vdots \\ 0 \\ 0 \\
\midrule
-b_1 \\ -b_2 \\ \vdots \\ -b_{\scriptscriptstyle m-1}
\end{array}\right]
}
\end{equation}
This linear system can be solved to obtain the LSS
solution $v_i$ and $\eta_i$.

\subsection{Derivation of the linear system via
the continuous optimization approach}
\label{s:kktdisc}
Problem (\ref{lsq2}) is constrained by a differential equation.
Its optimality condition must be derived using calculus of variation.
Denote $w(t)$ as the Lagrange multiplier
function; the Lagrangian of Problem (\ref{lsq2}) is
\[
\Lambda=
   \int_{0}^{T} \left(v^{\top} v + \alpha^2
   \eta^2 + 2\,w^{\top}\left(
   \frac{dv}{dt} - \frac{\partial f}{\partial u} v
            - \frac{\partial f}{\partial s} - \eta f\right)\right)\,dt
\]
The optimality condition requires a zero variation of
$\Lambda$ with respect to arbitrary $\delta w$, $\delta v$ and
$\delta\eta$.  This condition, through integration by parts,
transforms into the following differential equations and
boundary conditions
\[\left\{\begin{aligned}
&\frac{dv}{dt} - \frac{\partial f}{\partial u} v - \frac{\partial
f}{\partial s} - \eta f = 0\\
&\frac{dw}{dt} + \frac{\partial f}{\partial u}^{\top} w - v = 0  \\
&w(0) = w(T) = 0 \\
&\alpha^2\eta - w{\top} f = 0 \;.\\
\end{aligned} \right.\]
These linear differential equations consistently discretize
into the same linear system (\ref{kkt}) derived in the last subsection.

\subsection{Solution of the linear system}
\label{s:solvekkt}
The KKT system (\ref{kkt}) can be solved by using Gauss elimination to
remove the lower-left
block, forming the Schur complement
\begin{equation} \label{SchurSys}
{\bf B} {\bf B}^T {\bf w} = {\bf b}\;,
\end{equation}
where
\begin{equation}\addtolength{\arraycolsep}{-1mm} \label{Bmat}
{\bf B} =
\left[\begin{array}{cccccccc}
E_1 & \frac{f_1}{\alpha} & G_1    &        &        &   &    \\
    & & E_2 & \frac{f_2}{\alpha} & G_2    &        &    &    \\
    & &  &    & \ddots & \ddots &      &       \\
    & &  &   &        & E_m& \frac{f_m}{\alpha} & G_m    
\end{array}\right]\;,\quad
{\bf w} = 
\left[\begin{array}{c} w_1 \\ w_2 \\ \vdots \\ w_m
\end{array}\right]\;,\quad
{\bf b} = 
\left[\begin{array}{c} b_1 \\ b_2 \\ \vdots \\ b_m
\end{array}\right]\,.
\end{equation}

This Schur complement matrix ${\bf B} {\bf
B}^T$ is symmetric-positive-definite and block-tri-diagonal;
its block size is the dimension of the
dynamical system $n$.  Equation (\ref{SchurSys})
can be solved using a banded direct solver
with $O(m\,n^3)$ floating point operations \cite{Golub1996}.
One can also apply sparse QR factorization to the block-bi-diagonal
${\bf B}^T$,
and then use backward and forward substitution to compute $\bf w$.
The factorization also takes $O(m\,n^3)$ floating point
operations \cite{Golub1996}.  Iterative methods can be used when
$n$ is large.

$\bf w$ is substituted into the upper blocks of Equation (\ref{kkt})
to compute $v_i$ and $\eta_i$.
These blocks can be written as
\begin{equation} \label{veta}
v_{i+\frac12} = -G_i^T w_i - E_{i+1}^T w_{i+1} \;, \;\; 0\le i<m\;;\qquad
\eta_i = -\frac{f_i^T w_i}{\alpha^2}\;,\;\; 0<i<m\;.
\end{equation}
with the notation $w_0 = w_{m+1} = 0$. 
The desired derivative is then computed by discretizing
Equation (\ref{climatesens}) into
\begin{equation}\label{sensdJds}
\frac{d\langle J\rangle}{ds} 
 \approx
  \frac{1}{m} \sum_{i=0}^{m-1}\left(
  \frac{\partial J(u_{i+\frac12}, s)}{\partial u}\, v_{i+\frac12} +
  \frac{\partial J(u_{i+\frac12}, s)}{\partial s}\right)
+ \frac{1}{m-1}\sum_{i=1}^{m-1} \eta_i \widetilde{J}_i
\end{equation}
where
\begin{equation} \label{gdef} \begin{split}
\widetilde{J}_i &= 
\frac{J(u_{i-\frac12},s) + J(u_{i+\frac12},s)}{2} -
\frac{1}{m}\sum_{i=0}^{m-1} J(u_{i+\frac12},s) \;, \quad i=1,\ldots,m-1
\end{split} \end{equation}

\subsection{Summary of the algorithm}
\label{s:algorithm}
\begin{enumerate}
\item Choose a small time step size $\Delta t$ and
sufficient number of time steps $m$.
\item Compute a solution to the equation (\ref{ode}) at
$u_i=u_r\big((i+\frac12)\Delta t\big), i=0,\ldots,m-1$.
\item Compute the vectors and matrices $E_i$, $f_i$, $G_i$ and $b_i$
as defined in Equations (\ref{efgdef}).
\item Form matrix ${\bf B}$.  Choose an $\alpha$ so that $f_i/\alpha$
is on the same order of magnitude as $E_i$ and $G_i$.
Solve Equation (\ref{SchurSys}) for $\bf w$.
\item Compute $v_i$ and $\eta_i$ from Equation (\ref{veta}).
\item Compute desired derivative using Equation (\ref{sensdJds}).
\end{enumerate}
The computational cost is $O(m\,n^3)$ if a direct
solver is used for Equation (\ref{SchurSys}), where $m$ is the
number of time steps and $n$ is the dimension of the dynamical system.

\subsection{Adjoint formulation of the sensitivity computation method}
\label{s:adjoint}
The discrete adjoint computes the same derivative as in
Equation (\ref{sensdJds}) by first solving the adjoint system
\begin{equation} \label{kktadj} \addtolength{\arraycolsep}{-1mm}
\scriptstyle
\left[\begin{array}{cccccccc|cccc}
I   &   &    &   &    &        &     &   &    E_1^T &        &        &     \\
    &\alpha^2 &    &   &    &        &     &   &    f_1^T &        &        &     \\
    & & I    &   &    &        &     &   &    G_1^T & E_2^T  &        &     \\
    & &      & \alpha^2 &    &        &     &   &          & f_2^T  &        &     \\
    & &      &   & I &        &     &   &          & G_2^T  &        &     \\
    & &      &   &    & \ddots &     &   &          &        & \ddots & E_{\scriptscriptstyle m-1}^T \\
    & &      &   &    &        & \alpha^2 &   &          &        & & f_{\scriptscriptstyle m-1}^T \\
    & &      &   &    &        &     & I &          &        & & G_{\scriptscriptstyle m-1}^T \\
\midrule
E_1 & f_1 & G_1    &        &        &   &    &       &        &        &     \\
    & & E_2 & f_2 & G_2    &        &    &   &       &        &        &     \\
    & &  &    & \ddots & \ddots &      & &     &        &        &     \\
    & &  &   &        & E_{\scriptscriptstyle m-1}& f_{\scriptscriptstyle m-1} & G_{\scriptscriptstyle m-1}    &       &        &        &     \\
\end{array}\right]
\left[\begin{array}{c}
\hat{v}_{\frac12} \\ \hat{\eta}_1 \\ \hat{v}_{1+\frac12} \\ \hat{\eta}_2
\\ \hat{v}_{2+\frac12} \\
\vdots \\ \hat{\eta}_{\scriptscriptstyle m-1} \\ \hat{v}_{m-\frac12} \\
\midrule \hat{w}_1 \\ \hat{w}_2 \\ \vdots \\ \hat{w}_{\scriptscriptstyle m-1}
\end{array}\right]
=
\left[\begin{array}{c}
\frac{1}{m}\frac{\partial J(u_{1/2}, s)}{\partial u} \\
\quad\frac{1}{m-1}\widetilde{J}_1 \\
\frac{1}{m}\frac{\partial J(u_{1+1/2}, s)}{\partial u} \\
\quad\frac{1}{m-1}\widetilde{J}_2  \\
\frac{1}{m}\frac{\partial J(u_{2+1/2}, s)}{\partial u} \\ \vdots \\
\quad\frac{1}{m-1}\widetilde{J}_{m-1}\\
\frac{1}{m}\frac{\partial J(u_{\scriptscriptstyle m-1/2}, s)}{\partial u} \\
\midrule
0 \\ 0 \\ \vdots \\ 0
\end{array}\right]
\end{equation}
The system has the same matrix as
Equation (\ref{kkt}), but a different right hand side.  It can
be solved by inverting
\begin{equation} \label{SchurSys2}
{\bf B} {\bf B}^T {\bf \hat{w}} = {\bf B}{\bf g}\;,
\end{equation}
where ${\bf B}$ is defined in Equation (\ref{Bmat}), ${\bf \hat{w}} =
(\hat{w}_1, \ldots, \hat{w}_{m-1})$, and $\bf g$ is the
upper part of Equation (\ref{kktadj})'s right hand side.
Once $\bf \hat{w}$ is computed, $d\langle J\rangle/ds$ can be computed via
\begin{equation}\label{adjdJds}
\frac{d\langle J\rangle}{ds} \approx 
 \sum_{i=1}^{m-1} b_i^T \hat{w}_i
 + \frac{1}{m} \sum_{i=0}^{m-1}\frac{\partial J(u_{i+\frac12}, s)}{\partial s}\;,
\end{equation}
where $b_i$ is defined in Equation (\ref{efgdef}).
This adjoint derivative
equals to the derivative computed in Section \ref{s:algorithm} up to
round-off error.  The examples in this paper use the
algorithm in Section \ref{s:algorithm}.

\section{Application to the Van der Pol oscillator}
\label{s:vdp}

We apply our method to the Van der Pol oscillator
\begin{equation} \label{vdp}
   \frac{d^2y}{dt^2} = -y + \beta (1 - y^2) \frac{dy}{dt}\;.
\end{equation}
\begin{figure}[htb] \centering
    \subfloat[Limit cycle attractors of the Van der Pol oscillator
    at $\beta=0.2,0.8,1.6$ and $2.0$.]
    {\label{f:vanderpolPhase}
    \includegraphics[width=2.2in,trim=0 0.0cm 0 .9cm,clip]{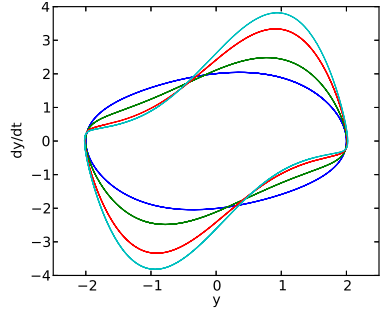}}
    \hspace{0.1in}
    \subfloat[For each value of $\beta$, $\langle J\rangle^{\frac18}$ is
    estimated 20 times by solving initial value problems of length $50$ with
    random initial conditions.]{
    \includegraphics[width=2.2in,trim=0 0.0cm 0 .9cm,clip]{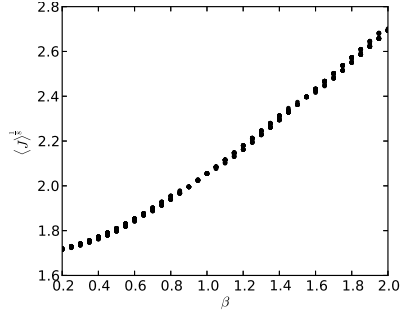}}\\
    \subfloat[$d\langle J\rangle^{\frac18}/d\beta$
    estimated by finite differencing pairs of trajectories
    with $\Delta\beta=0.05$.  For each value of $\beta$, the black dots
    are computed on 20 pairs of trajectories with length $50$.
    The red line is computed on pairs of trajectories with length
    $5000$.]{
    \includegraphics[width=2.2in,trim=0 0.0cm 0 .9cm,clip]{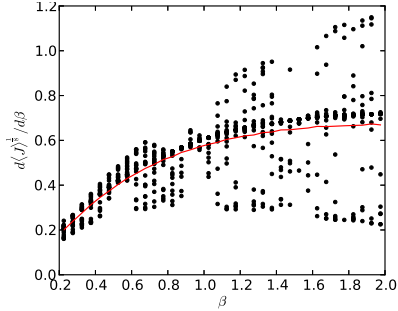}}
    \hspace{0.1in}
    \subfloat[$d\langle J\rangle^{\frac18}/d\beta$ estimated
    with Least Squares Shadowing sensitivity analysis.
    For each value of $\beta$, the black dots
    are computed on 20 trajectories of length $50$.
    The red line is computed on trajectories of length
    $5000$.]{
    \includegraphics[width=2.2in,trim=0 0.0cm 0 .9cm,clip]{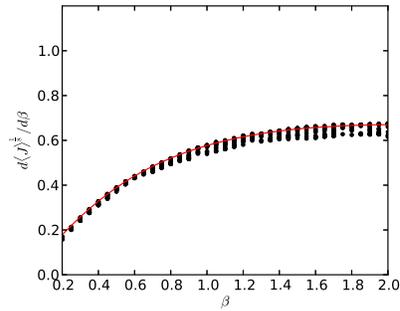}}
\caption{Least Squares Shadowing Sensitivity Analysis
of the van der Pol oscillator.}
\label{f:vdp}
\end{figure}
to compute sensitivity to the parameter $\beta$ in the system.
Figure \ref{f:vanderpolPhase}
shows the limit cycle attractor as $\beta$ varies from
$0.2$ to $2.0$.   As $\beta$ increases,
the maximum magnitude of $dy/dt$ significantly increases.  We choose the
objective function to be the $L^8$ norm of $dy/dt$, which has a
similar trend to the $L^{\infty}$ norm and reflects the magnitude of the
peak in $dy/dt$.
By denoting $u=(u^{(1)}, u^{(2)})= (y, dy/dt)$ as the state vector, we convert
the second order ODE (\ref{vdp}) into two coupled first order ODEs, and
write the objective function as
\begin{equation}
   \langle J\rangle^{\frac18} = \left( \lim_{T\rightarrow\infty}\frac1T \int_0^T
       J(u, \beta) \,dt \right)^{\frac18}\;,\quad
   J(u, \beta) = \left(u^{(2)}\right)^8
\end{equation}

The method described in Section \ref{s:method} is then applied to
compute $v$:  for each $\beta$, we start the simulation by assigning
uniform $[0,1]$ random numbers to $(u^{(1)}, u^{(2)})$ as their initial
condition at $t=-50$.  This initial time is chosen to be large enough
so that when the ODE is integrated to $t=0$, its state $u(0)$
is on its attractor.
A trajectory $u(t), 0\le t\le 50$ is then computed
using a scipy\cite{scipy} wrapper of lsoda\cite{lsoda},
with time step size
$\Delta t=0.02$.  The trajectory is about 50 times
the longest timescale of the system.
The $m=2500$ states along the resulting trajectory
are used to construct the coefficient in Equation (\ref{kkt}).

The solution to Equation (\ref{kkt}) is then substituted into Equation
(\ref{sensdJds}) to estimate the derivative
of the $\langle J\rangle$ to the parameter $\beta$.  Finally, the derivative
of the output $\langle J\rangle^{\frac18}$ is computed using
\begin{equation} \label{dJvanderpol}
   \frac{d\langle J\rangle^{\frac18}}{d\beta} = 
   \frac{\langle J\rangle^{-\frac78}}{8} \frac{d\langle J\rangle}{d\beta}\;.  \end{equation}
The computed derivative is compared against finite difference in
Figure \ref{f:vdp}.  For each value of $\beta$, we repeat both the finite
difference and least squares shadowing 20 times on randomly initialized
trajectories; the spread of the computed derivatives
represents the approximation error due to
insufficient trajectory length.  Long trajectories are used to compute
more accurate derivatives.
The results indicate that the least squares shadowing method is
more accurate than finite difference in this problem
with the same trajectory length.

\section{Application to the Lorenz system}
\label{s:lorenz}

\begin{figure}[htb!] \centering
    \subfloat[Attractors of the Lorenz system at $\rho=10$ (open circle),
    $\rho=25,50,75$ and $100$ (blue, green, red and black lines,
    respectively)]{ \label{f:lorenzPhase}
    \includegraphics[width=2.2in,trim=0 0 0 0,clip]{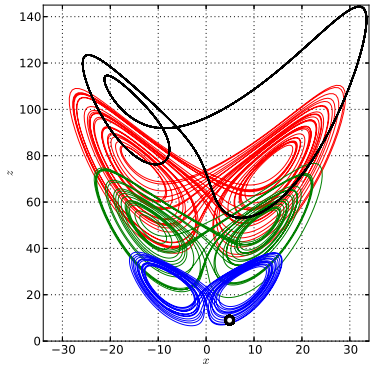}}
    \hspace{0.1in}
    \subfloat[ For each value of $\rho$, $\langle z\rangle$ is
    estimated 20 times by solving initial value problems of length $50$ with
    random initial conditions.]{
    \includegraphics[width=2.2in,trim=0 0 0 0,clip]{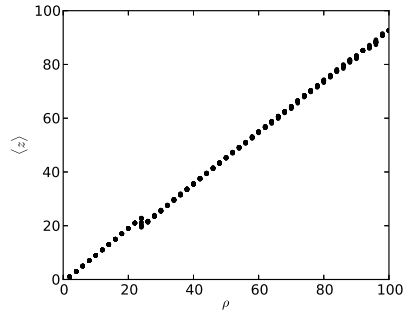}}\\
    \subfloat[$d\langle z\rangle/d\rho$
    estimated by finite differencing pairs of trajectories
    with $\Delta\rho=2$.  For each value of $\rho$, the black dots
    are computed on 20 pairs of trajectories with length $50$.
    The red line is computed on pairs of trajectories with length
    $5000$.]{
    \includegraphics[width=2.2in,trim=0 0 0 0,clip]{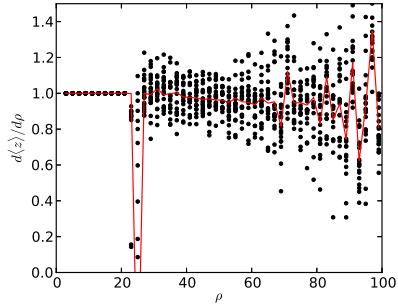}}
    \hspace{0.1in}
    \subfloat[$d\langle z\rangle/d\rho$ estimated
    with Least Squares Shadowing sensitivity analysis.
    For each value of $\rho$, the black dots
    are computed on 20 trajectories of length $50$.
    The red line is computed on trajectories of length
    $5000$.]{
    \includegraphics[width=2.2in,trim=0 0 0 0,clip]{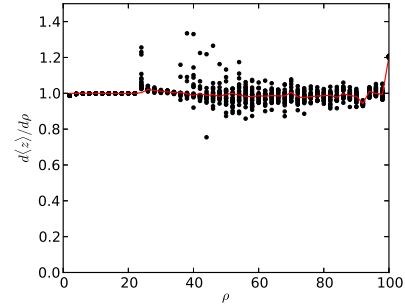}}
\caption{Least Squares Shadowing Sensitivity Analysis of the Lorenz system.}
\label{f:lorenz}
\end{figure}

We apply our method to the Lorenz system
\begin{equation} \label{lorenz}
\frac{dx}{dt} = \sigma(y-x)\;,\quad
\frac{dy}{dt} = x(r-z) - y\;,\quad
\frac{dz}{dt} = xy - \beta z\;.
\end{equation}
and analyze sensitivity to the parameter $\rho$ in the system.
The behavior of the Lorenz system as $\rho$ changes from $0$ to $100$ is
shown in Figure \ref{f:lorenzPhase}, and can be summarized as following
\cite{lorenzbook}:
\begin{itemize}
\item Stable fixed point attractor at $(0,0,0)$ for $0\le \rho<=1$.
\item Two stable fixed point attractors at
$x=y=\pm\sqrt{\beta(\rho-1)},z=\rho-1$ for $1<\rho<24.74$.
\item Quasi-hyperbolic strange attractors for $24.06<\rho<31$.  This
includes the classic Lorenz attractor at $\rho=28$.
\item Non-hyperbolic quasi-attractors for $31<\rho<99.5$.
\item Periodic limit cycle attractors with
an infinite series of period doubling for $\rho>99.5$.
\end{itemize}
Despite the many transitions in the fundamental nature of the system,
the mean $z$ value
\begin{equation}
   \langle z\rangle = \lim_{T\rightarrow\infty}\frac1T \int_0^T z \,dt
\end{equation}
apparently increases as the parameter $\rho$ increases.  $\langle
z\rangle$ is chosen to be our time averaged output quantity in this study.

By denoting $u= (x,y,z)$,
the method described in Section \ref{s:algorithm} is applied to the
Lorenz system.  For each $\rho$, we start the simulation at $t=-50$
with uniform $[0,1]$ random numbers as initial conditions for $x,y$ and $z$.
The Lorenz system is
integrated to $t=0$, so that $u(0)$ is approximately on the
attractor.  A trajectory $u(t), 0\le t\le 50$ is then computed
using a scipy\cite{scipy} wrapper of lsoda\cite{lsoda},
with time step size
$\Delta t=0.01$.  The resulting $m=5000$ states along the trajectory
are used to construct the linear system (\ref{kkt}), whose solution
is then used to estimate the desired derivative
$d\langle z\rangle/d\rho$ using Equation (\ref{climatesens}).

\begin{figure}[htb!] \centering
    \subfloat[For each time length $T$, the Least squares shadowing
    algorithm runs on 10 random trajectories, computing 10 different
    derivatives.]{
    \includegraphics[width=2.2in,trim=0 0 0 0,clip]{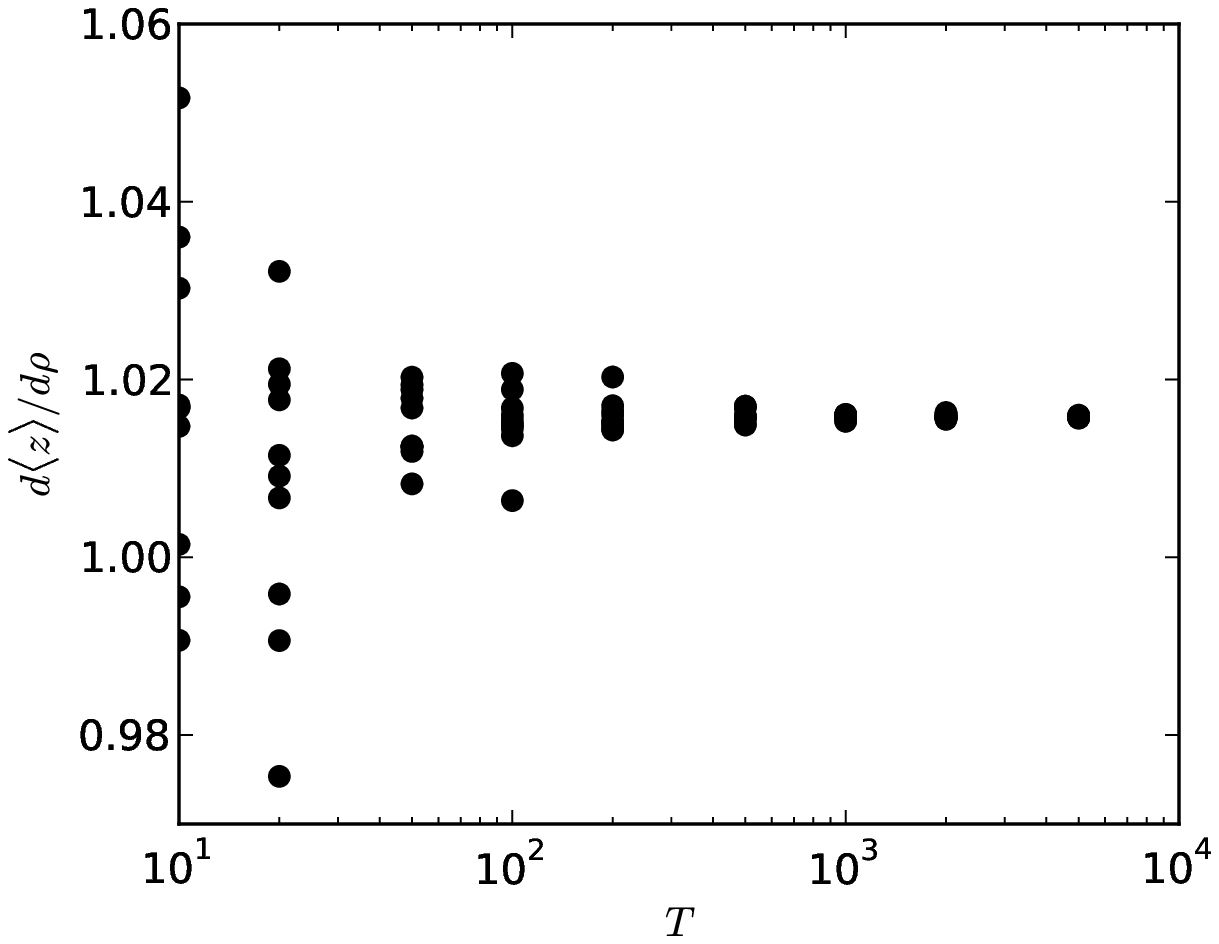}}
    \hspace{0.1in}
    \subfloat[The sample standard deviation of the 10 derivatives at
    each trajectory length $T$.]{
    \includegraphics[width=2.2in,trim=0 0 0 0,clip]{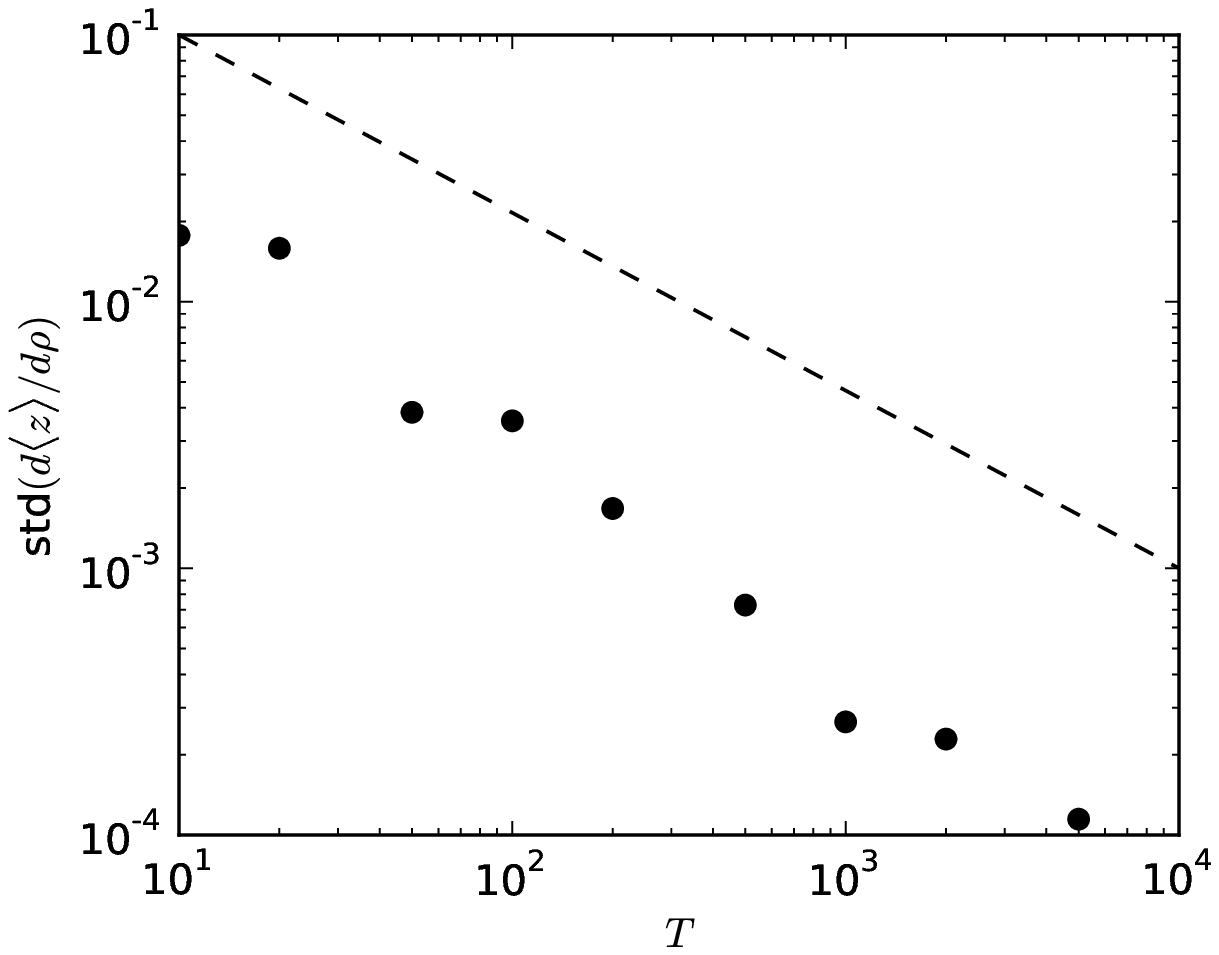}}
\caption{Convergence of Least Squares Shadowing Sensitivity Analysis
applied to the Lorenz system.}
\label{f:lorenzconv}
\end{figure}
The computed derivative is compared against finite difference values in
Figure \ref{f:lorenz}.  The dip in the finite difference value
at around $\rho=22.5$ is due to a bifurcation from fixed
point attractors to strange attractors at $24.0\le \rho\le 24.74$
(the two types of attractors co-exist within this range).
For $24.74<\rho<31$, the Lorenz system is dominated by a
quasi-hyperbolic attractor.  Least squares shadowing
sensitivity analysis computes accurate and consistent gradients on
randomly chosen short trajectories on the attractor.  The
computed gradients has a random error on the order of $O(T^{-\frac12})$,
a result derived theoretically for discrete-time dynamical systems
\cite{lsstheory} and shown empirically here in Figure \ref{f:lorenzconv}.

As $\rho$ increases beyond $31$, the system is non-hyperbolic and
its trajectories form an object known as a quasi-attractor
\cite{bonatti2010dynamics}.
For $\rho>99.5$, the system
transitions to periodic oscillations, then goes through an infinite
series of period doubling bifurcations.
Despite of the complex, non-hyperbolic behavior, our method computes
derivatives that are more accurate than finite difference on the
same trajectory lengths.

\section{Application to an aero-elastic limit cycle oscillator}
\label{s:aeroelastic}

\begin{figure}[htb!] \centering
    \includegraphics[width=3.2in]{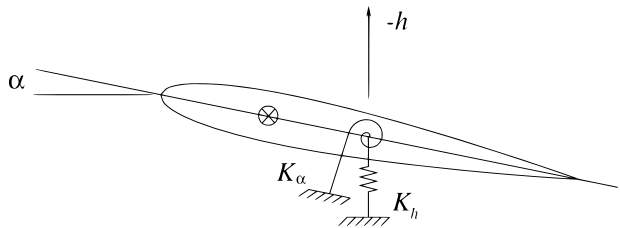}
\caption{Model aero-elastic oscillator}
\label{f:airfoil}
\end{figure}

We apply our method to a simple model of aeroelastic limit cycle
oscillation, as shown in Figure \ref{f:airfoil}.
The model is described in detail by Zhao and Yang\cite{chaotic_lco}.
The governing equations are
\begin{equation}\label{stiff}
\begin{split}
    &\frac{d^2h}{dt^2} + 0.25\,\frac{d\alpha}{dt} + 0.1\, \frac{dh}{dt} + 0.2\, h
    + 0.1\, Q\, \alpha = 0 \\
    &0.25\, \frac{d^2h}{dt^2} + 0.5\, \frac{d^2\alpha}{dt^2} + 0.1\, \frac{d\alpha}{dt}
    + 0.5\, \alpha + 20\, \alpha^3 - 0.1\, Q\, \alpha = 0
\end{split}
\end{equation}
where $h$ is the plunging degree of freedom, and $\alpha$ is the
pitching degree of freedom.
We analyze sensitivity to the reduced dynamic pressure $Q$.

\begin{figure}[htb!] \centering
    \subfloat[Bifurcation diagram in the parameter range considered.]{
    \label{f:stiffbif}
    \includegraphics[width=2.2in,trim=0 0 0 0,clip]{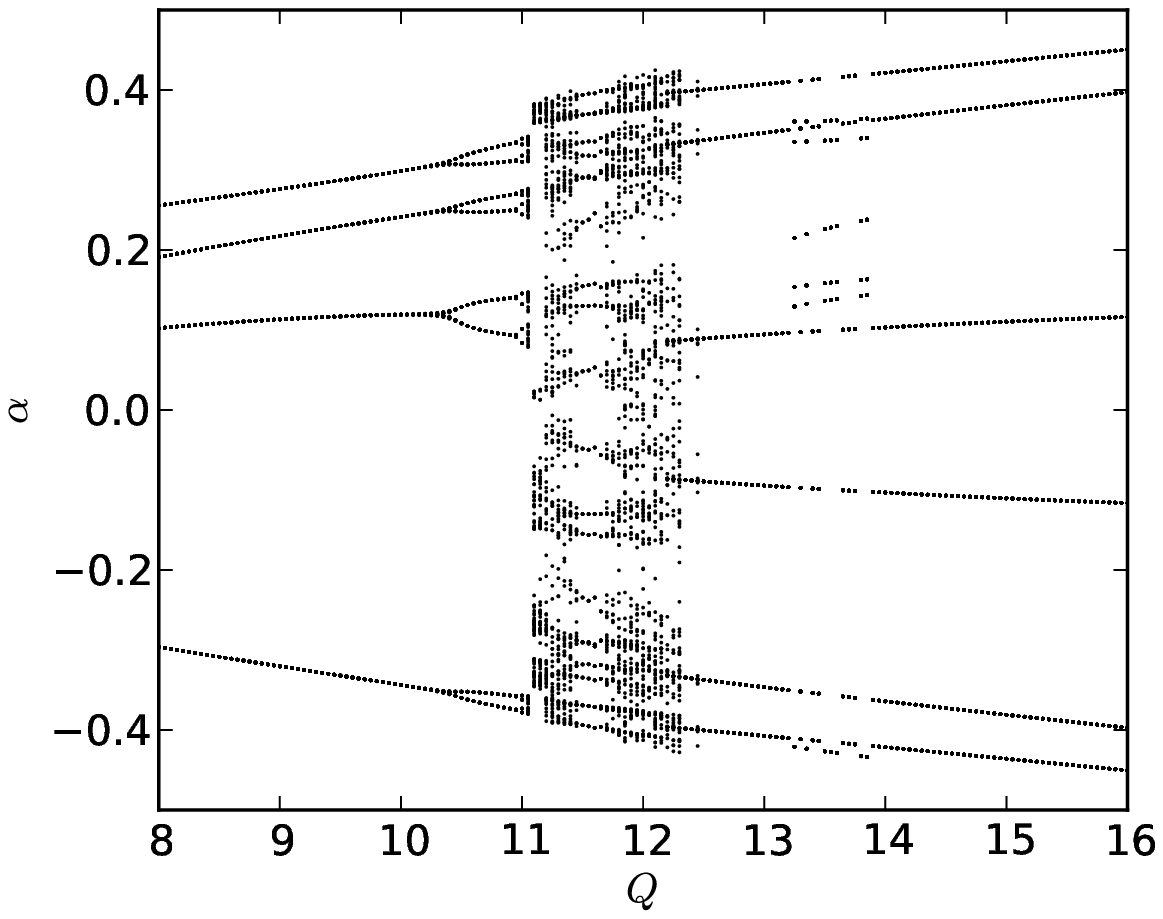}}
    \hspace{0.1in}
    \subfloat[Phase plots ($\alpha$ vs $\dot\alpha=d\alpha/dt$)
    at $Q=8$ (black), $Q=12$ (green) and $Q=16$ (red).]{\label{f:stiffphase}
    \includegraphics[width=2.2in,trim=0 0 0 0,clip]{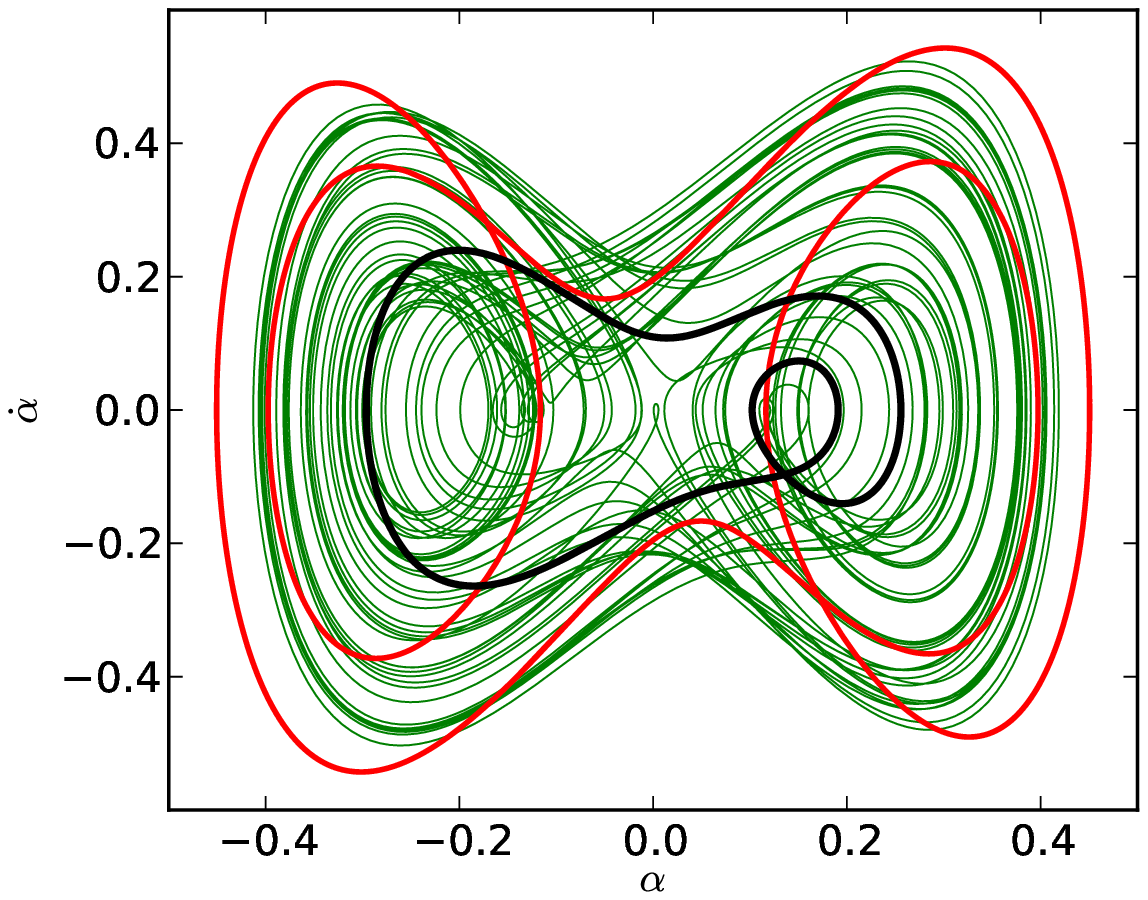}}\\
    \subfloat[$d\langle J\rangle^{\frac18}/dQ$
    estimated by finite differencing pairs of trajectories
    with $\Delta Q=0.2$.  For each value of $Q$, the black dots
    are computed on 20 pairs of trajectories with length $300$.]{
    \includegraphics[width=2.2in,trim=0 0 0 0,clip]{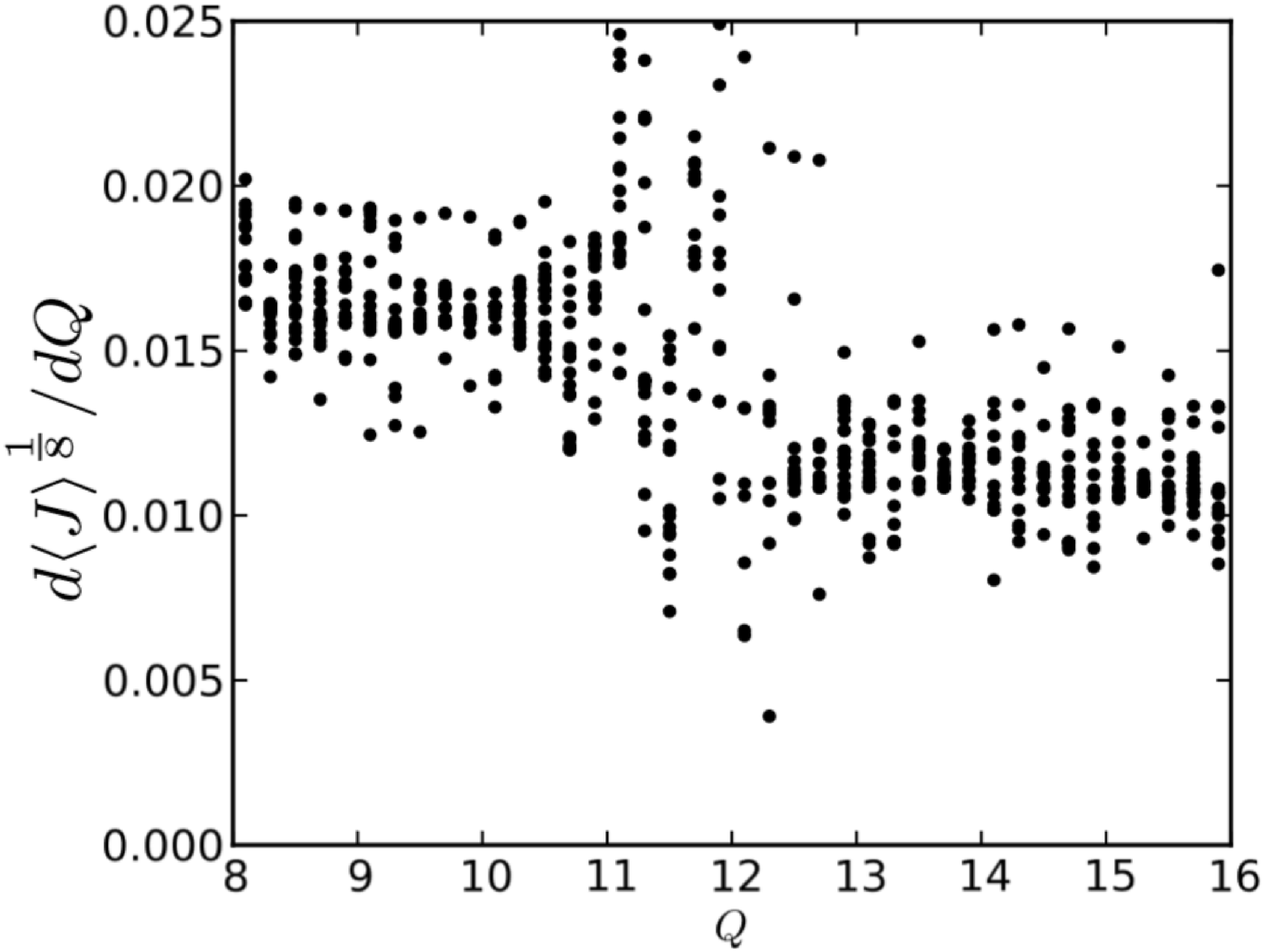}}
    \hspace{0.1in}
    \subfloat[$d\langle J\rangle^{\frac18}/dQ$ estimated
    with Least Squares Shadowing sensitivity analysis.
    For each value of $Q$, the black dots
    are computed on 20 trajectories of length $300$.
    The red line is computed on trajectories of length
    $30000$.]{
    \includegraphics[width=2.2in,trim=0 0 0 0,clip]{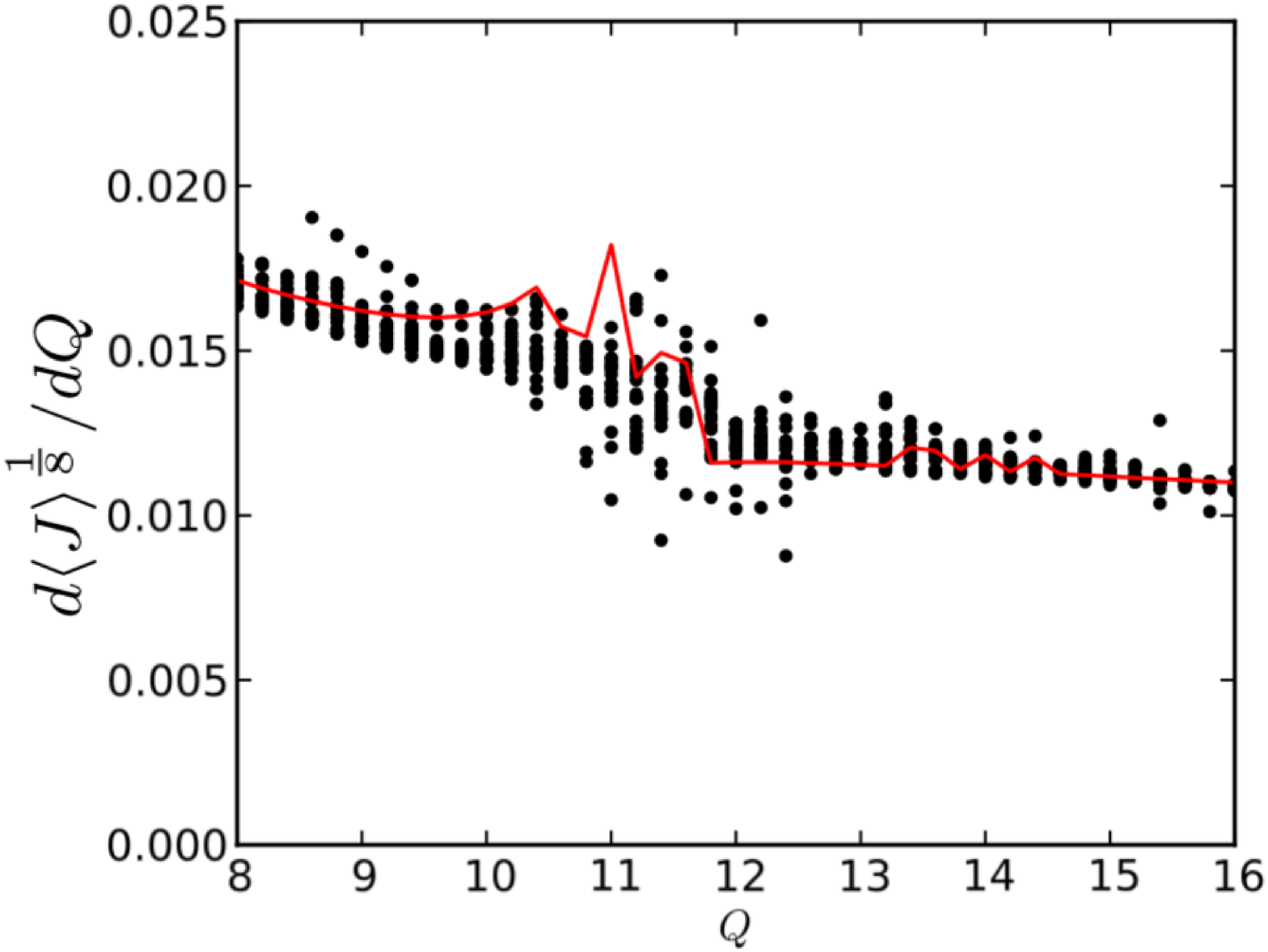}}
\caption{Least Squares Shadowing Sensitivity Analysis on the
aero-elastic oscillator model (\ref{stiff}).}
\label{f:stiff}
\end{figure}

The bifurcation diagram of $\alpha$ as $Q$ increases
from 8 to 16 is shown in Figure \ref{f:stiffbif}.
The behavior of the system as $Q$ varies is complex
\cite{Lee_Price_Wong_1999}:
At low values of $Q$, the system has an asymmetric limit cycle
attractor.  As $Q$ increases beyond about 10.25, a series of period
doubling bifurcations occurs, leading to transition into chaos just
beyond $Q=11$.  At about $Q=12.5$, the system ceases to be chaotic, and
transitions to symmetric periodic limit cycle oscillation.
When $Q$ increases beyond about $13.25$, there appears to be
small windows of asymmetric oscillations.
Finally, at about $Q=13.9$, the system recovers symmetric
periodic limit cycle oscillations.
The phase plot of the system at several values of $Q$ is shown in Figure
\ref{f:stiffphase}.  These include an asymmetric periodic limit cycle
attractor at $Q=8$, a chaotic limit cycle attractor or quasi-attractor
at $Q=12$, and a symmetric periodic limit cycle attractor at $Q=16$.

We observe that the magnitude of the
oscillation grows as $Q$ increases, and choose the $L^8$ norm of
the pitch angle $\alpha$ as the objective function.  The $L^8$ norm
has similar trend as the $L^{\infty}$ norm, and indicates
the magnitude of the oscillation in the pitching degree of freedom.
Denoting $u=(u^{(1)},u^{(2)},u^{(3)},u^{(4)})
= (y, \alpha, dy/dt, d\alpha/dt)$ as the state vector,
we convert the pair of second order ODEs (\ref{stiff}) into a system of four
first order ODEs.  The output can then be written as
\begin{equation}
   \langle J\rangle^{\frac18} = \left( \lim_{T\rightarrow\infty}\frac1T
   \int_0^T u^{(2)\,8} \,dt \right)^{\frac18}
\end{equation}
We use the method described in Section \ref{s:algorithm} to compute the
derivative of the objective function to the input parameter
$Q$.  For each $Q$, we initiate the simulation at $t=-300$
with uniform $[0,1]$ random numbers as its initial condition.
The ODE is integrated to $t=0$ to ensure that $u(0)$ is approximately on
an attractor.  A trajectory $u(t), 0\le t\le 300$ is then computed
using a scipy\cite{scipy} wrapper of lsoda\cite{lsoda},
with time step size
$\Delta t=0.02$.  The resulting $15000$ states along the trajectory
are used to construct the linear system (\ref{kkt}), whose solution is
used to estimate the derivative
of the output with respect to $Q$.
The computed derivative is compared against finite difference values in
Figure \ref{f:stiff}.
Whether the system exhibits periodic or chaotic limit cycle oscillations,
the derivative computed using least
squares shadowing sensitivity analysis is more accurate than finite
difference results.

\section{Conclusion}
\label{s:conclude}

We presented the Least Squares Shadowing method for computing
derivatives in ergodic dynamical systems.
Traditional tangent and adjoint methods linearize the 
ill-conditioned initial value problem, thereby computing
large derivatives useless for control, optimization and inference
problems.
The new method linearizes the well-conditioned least squares shadowing
problem, thereby computing useful derivatives of long time averaged
quantities.  The method is
demonstrated on the periodic van der Pol oscillator, the chaotic Lorenz
attractor, and a simple aero-elastic oscillation model that exhibits
mixed periodic and chaotic behavior.  These applications demonstrate
the effectiveness of our new
sensitivity computation algorithm in many complex nonlinear dynamics
regimes.  These include fixed points, limit cycles, quasi-hyperbolic and
non-hyperbolic strange attractors.

The Least Squares Shadowing method requires solving either a sparse matrix
system (in its discrete formulation) or a boundary value problem in time
(in its continuous formulation).  This boundary value problem is
about twice as large as a linearized initial value problem, in terms of
the dimension and sparsity of the matrix for the discrete formulation,
and in terms of the number of equations for the continuous formulation.
When the dynamical system is low dimensional,
the sparse matrix system can be solved using a direct matrix
solver; computing the derivative of the output costs a few times
more than computing the output itself by solving an initial value problem.
When the dynamical system is high dimensional, e.g., a discretized
partial differential equation, iterative solution methods should be
used instead of direct matrix solvers.
Because the system is well-conditioned and only twice as large as an
initial value problem, an iterative solution can potentially cost only a small
multiple of an initial value solution, particularly if using an iterative
solver specifically designed for this problem.
Therefore, we think that the Least Squares Shadowing method
is not only efficient for low-dimensional
chaotic dynamical systems, but also applicable to sensitivity analysis of
large chaotic dynamical systems.

\section*{Acknowledgments}
The first author acknowledges AFOSR Award F11B-T06-0007
under Dr. Fariba Fahroo, NASA Award NNH11ZEA001N under Dr. Harold Atkins,
and a subcontract of the DOE PSAAP Program at Stanford.

\bibliography{master}

\appendix
\section{Derivation of Equations (\ref{lsq2}) and (\ref{climatesens})}
If $\frac{du_r}{dt} = f(u_r,s)$ in Problem (\ref{lsq0}), then
$u_{lss}(t;s)\equiv u_r(t)$ and $\tau_{lss}(t;s)\equiv t$ solve
the problem.
Because Problem (\ref{lsq0}) is well-conditioned, its solution
at a perturbed parameter value $s+\delta s$ for the same $u_r$ should be
slightly different.  Denote
\begin{equation}
v(t) := \frac{d}{ds}\Big(u_{lss}(\tau_{lss}(t;s);s) - u_r(t)\Big)\;,\quad
\eta(t) := \frac{d}{ds} \left(\frac{d\tau_{lss}(t;s)}{dt}-1\right)\;,
\end{equation}
which for infinitesimal $\delta s$ translate into
\begin{equation} \begin{aligned}
& \tau_{lss}(t;s+\delta s) = \int_0^t (1 + \eta(t') \delta s)\, dt' \;,\\
& u_{lss}(\tau_{lss}(t;s+\delta s);s+\delta s) = u_r(t) + v(t) \delta s\;.
\end{aligned} \end{equation}
The second equation translates the objective function in Problem
(\ref{lsq0}) into the objective function in Problem (\ref{lsq2}).
$du_{lss}(t;s+\delta s)$ must satisfy the constraint in Problem
(\ref{lsq0}), which translates into (ignoring $O(\delta s^2)$ terms)
\begin{equation} \begin{aligned}
 & \frac{d}{dt} \big(u_r(t) + v(t) \delta s\big) \\
=& \frac{du_{lss}(\tau_{lss}(t;s+\delta s);s+\delta s)}{dt} \\
=& \frac{d\tau_{lss}(t;s+\delta s)}{dt}
   \frac{du_{lss}(\tau;s+\delta s)}{d\tau}\bigg|_{\tau=\tau_{lss}(t;s+\delta s)}\\
=& \frac{d\tau_{lss}(t;s+\delta s)}{dt}
f\Big(u_{lss}\big(\tau_{lss}(t;s+\delta s);s+\delta s\big), s+\delta s\Big) \\
=& (1 + \eta(t)\delta s) \left(f(u_r(t),s)
+ \frac{\partial f}{\partial u} v(t) \delta s
+ \frac{\partial f}{\partial s} \delta s \right) \\
=& f(u_r(t),s) + \eta(t) f(u_r(t),s) \delta s
+ \frac{\partial f}{\partial u} v(t) \delta s
+ \frac{\partial f}{\partial s} \delta s \\
\end{aligned} \end{equation}
Because $\frac{du_r}{dt} = f(u_r,s)$,
we cancel all $O(1)$ terms, leaving only
\begin{equation} \begin{aligned}
   \frac{dv}{dt}
= \eta(t) f(u_r(t),s) + \frac{\partial f}{\partial u} v(t)
 + \frac{\partial f}{\partial s}\;,
\end{aligned} \end{equation}
the constraint in the linearized least squares shadowing problem
(\ref{lsq2}).

For infinitesimal $\delta s$, the
definition of $\overline{J}^{(T)}_{lss}(s)$ in Equation 
(\ref{lssobj}) leads to
\begin{equation}\begin{aligned}
 &\overline{J}^{(T)}_{lss}(s+\delta s) - \overline{J}^{(T)}_{lss}(s) \\
=& \frac{\displaystyle\int_{\tau(0;s+\delta s)}^{\tau(T;s+\delta s)}
         J(u_{lss}(t;s+\delta s), s+\delta s) \,dt}
   {\tau(T;s+\delta s)-\tau(0;s+\delta s)}
 - \frac{\displaystyle\int_{\tau(0;s)}^{\tau(T;s)}J(u_{lss}(t;s),s) \,dt}
        {\tau(T;s)-\tau(0;s)} \\
=& \frac{\displaystyle\int_{0}^{T} J(u_{lss}(\tau_{lss}(t;s+\delta s),
    s+\delta s), s+\delta s) \frac{d\tau(s;s+\delta s)}{dt} \,dt}
   {\tau(T;s+\delta s)-\tau(0;s+\delta s)}
 - \frac{\displaystyle\int_{0}^{T} J(u_r(t),s) \,dt}{\tau(T;s)-\tau(0;s)} \\
=& \frac{\displaystyle\int_{0}^{T} J(u_{lss}(\tau_{lss}(t;s+\delta s),
    s+\delta s), s+\delta s) \frac{d\tau(s;s+\delta s)}{dt} \,dt}
   {\displaystyle\int_0^T(1+\eta(t')\delta s)dt'}
 - \frac{\displaystyle\int_{0}^{T} J(u_r(t),s) \,dt}
   {\displaystyle\int_0^T(1+\eta(t')\delta s)dt'} \\
&+ \frac{\displaystyle\int_{0}^{T} J(u_r(t),s) \,dt}
   {\displaystyle\int_0^T(1+\eta(t')\delta s)dt'} 
 - \frac{\displaystyle\int_{0}^{T} J(u_r(t),s) \,dt}{T} \\
=& \frac{\displaystyle\int_{0}^{T} \left(\left(J(u_r(t),s) + \frac{\partial
J}{\partial u} v(t)\,\delta s + \frac{\partial J}{\partial s}\delta
s\right)\big( 1 + \eta(t)\big)
    - J(u_r(t),s)\right) dt}{\displaystyle\int_0^T(1+\eta(t')\delta s)dt'} \\
&+ \left(\int_{0}^{T} J(u_r(t),s) \,dt\right)
   \frac{-\displaystyle\int_0^T \eta(t') \delta s \,dt'}
   {T\displaystyle\int_0^T(1+\eta(t')\delta s)dt'} \\
=& \left(\frac{\displaystyle\int_{0}^{T} \left(\frac{\partial
J}{\partial u} v(t) + \frac{\partial J}{\partial s} + \eta(t) J(u_r(t),s)
    \right) dt}{\displaystyle\int_0^T(1+\eta(t')\delta s)dt'}  \right.\\
&- \left.\frac{\left(\displaystyle\int_{0}^{T} J(u_r(t),s) \,dt\right)
   \left(\displaystyle\int_0^T \eta(t') \,dt'\right)}{T^2}\right)
   \delta s + O(\delta s^2) \\
=& \frac{\delta s}T\int_{0}^{T} \left(\frac{\partial
J}{\partial u} v(t) + \frac{\partial J}{\partial s} + \eta(t)
  \Big( J(u_r(t),s) - \overline{J}^{(T)}_{lss} \Big) \right) dt
    + O(\delta s^2) \\
\end{aligned}\end{equation}
Therefore,
\begin{equation}\begin{aligned}
 \frac{d\overline{J}^{(T)}_{lss}}{ds}
&= \lim_{\delta s\to0}\frac{\overline{J}^{(T)}_{lss}(s+\delta s) - \overline{J}^{(T)}_{lss}(s)}{\delta s} \\
&= \frac1T\int_{0}^{T} \left(\frac{\partial
J}{\partial u} v(t) + \frac{\partial J}{\partial s} + \eta(t)
  \Big( J(u_r(t),s) - \overline{J}^{(T)}_{lss} \Big) \right) dt
\end{aligned}\end{equation}

\end{document}